\documentstyle[preprint,aps,epsfig]{revtex}

\newcommand{\lbar}{\overline}

\newcommand{\beq}{\begin{equation}}
\newcommand{\eeq}{\end{equation}}
\newcommand{\beqs}{\begin{eqnarray}}
\newcommand{\eeqs}{\end{eqnarray}}

\newcommand{\BR}{\mbox{{\rm BR}}}

\newcommand{\MuToEee}{\mu^- \to e^- \, e^+ \, e^-}
\newcommand{\TauToEee}{\tau^- \to e^- \,  e^+ \,  e^-}
\newcommand{\MuToEonTi}{\sigma(\mu^- \, \text{Ti} \to e^- \, \text{Ti}) \over
                        \sigma(\mu^- \, \text{Ti} \to \mbox{capture})}
\newcommand{\MuToEonPb}{\sigma(\mu^- \, \text{Pb} \to e^- \, \text{Pb}) \over
                        \sigma(\mu^- \, \text{Pb} \to \mbox{capture})}
\newcommand{\MuToEonS}
 {\sigma(\mu^- \, ^{32}\text{S} \to e^- \, ^{32}\text{S}) \over
  \sigma(\mu^- \, ^{32}\text{S} \to \nu_\mu \, ^{32}\text{P}^*)}

\newcommand{\TauToERho}{\tau^- \to e^- \, \rho^0}
\newcommand{\TauToEPi} {\tau^- \to e^- \, \pi^0}
\newcommand{\TauToEEta}{\tau^- \to e^- \, \eta}
\newcommand{\TauToEPiPi}{\tau^- \to e^- \, \pi^+ \, \pi^-}

\newcommand{\NSNI}{non-standard neutrino interactions}
\newcommand{\andthis}{~~~~~~~\mbox{and}~~~~~~~} 
\newcommand{\CL}{CL}   
\newcommand{\DOF}{DOF} 
\newcommand{\calS}{{\cal S}}
\newcommand{\calV}{{\cal V}}

\def\lsim{\ \rlap{\raise 3pt \hbox{$<$}}{\lower 3pt \hbox{$\sim$}}\ }
\def\gsim{\ \rlap{\raise 3pt \hbox{$>$}}{\lower 3pt \hbox{$\sim$}}\ }

\def\npb#1{Nucl.\ Phys.\ {\bf B\,#1}}
\def\npbps#1{Nucl.\ Phys.\ (Proc. Suppl.) {\bf B\,#1}}
\def\plb#1{Phys.\ Lett.\ {\bf B\,#1}}
\def\prc#1{Phys.\ Rev.\ {\bf C\,#1}}
\def\prd#1{Phys.\ Rev.\ {\bf D\,#1}}
\def\prl#1{Phys.\ Rev.\ Lett. {\bf#1}}

\def\epjc#1{Eur.~Phys.~J.\ {\bf C\,#1}}

\def\rmp#1{Rev.\ Mod.\ Phys.\ {\bf #1}}

\def\mpla#1{Mod. Phys. Lett. {\bf A\,#1}}
\def\sjnp#1{Sov. J. Nucl. Phys. {\bf #1}}
\def\JHEP#1{JHEP {\bf #1}}

\begin{document}
\textheight = 23.5cm

\draft
{\tighten 
\preprint{\vbox{\hbox{WIS-5/00/Apr-DPP}
                \hbox{MAD-NT/00-04}
                \hbox{hep-ph/0004049}
                \hbox{April 2000}}}
\title{~ \\ Status of the solution to the solar neutrino problem \\ 
            based on non-standard neutrino interactions}
\author{ 
S.\ Bergmann$^1$,
M.\ M.\ Guzzo$^{2}$, 
P.\ C.\ de Holanda$^{2}$, \\
P.\ I.\ Krastev$^{3}$ 
and 
H.\ Nunokawa$^{2}$ }
\address{\sl ~ \\
$^1$ Department of Particle Physics \\
     Weizmann Institute of Science, Rehovot 76100, Israel \\
\vspace{3mm}
$^2$ Instituto de F\' {\i}sica Gleb Wataghin \\
     Universidade Estadual de Campinas, UNICAMP \\    
     13083-970 Campinas SP, Brazil \\ 
\vspace{3mm}
$^3$ Department of Physics \\
     University of Wisconsin, Madison, WI 53706, USA}
\maketitle
\vspace{.5cm}
            
\hfuzz=25pt

\begin{abstract}

\noindent
We analyze the current status of the solution to the solar neutrino
problem based both on: a) non-standard flavor changing neutrino
interactions (FCNI) and b) non-universal flavor diagonal neutrino
interactions (FDNI). We find that FCNI and FDNI with matter in the sun
as well as in the earth provide a good fit not only to the total rate
measured by all solar neutrino experiments but also to the day-night
and seasonal variations of the event rate, as well as the recoil
electron energy spectrum measured by the SuperKamiokande
collaboration. This solution does not require massive neutrinos and
neutrino mixing in vacuum. Stringent experimental constraints on FCNI
from bounds on lepton flavor violating decays and on FDNI from limits
on lepton universality violation rule out $\nu_e \to \nu_\mu$
transitions induced by New Physics as a solution to the solar neutrino
problem. However, a solution involving $\nu_e \to \nu_\tau$
transitions is viable and could be tested independently by the
upcoming $B$-factories if flavor violating tau decays would be
observed at a rate close to the present upper bounds.

\end{abstract}

} 

\section{Introduction}

In recent years the accuracy with which the solar neutrino flux is
being measured has been improved
significantly~\cite{HS,GALLEX,SAGE,KK,SK}. Better statistics and
calibration of the pioneering experiments, as well as the first
next-generation experiment SuperKamiokande, measuring the solar
neutrino spectrum and the event rate as a function of the zenith angle
with unprecedented precision, have provided a lot of new information
about the solar neutrino problem~\cite{JNB}. On the theoretical side
several substantial improvements have been made in the standard solar
model (SSM)~\cite{BP95,BPBC,BBP,olderSSM} which now includes diffusion
of helium and heavy elements and updated low energy nuclear cross
sections relevant to the solar neutrino production~\cite{Adel}.
Furthermore, the SSM has received an important independent
confirmation by the excellent agreement between its predicted sound
speeds and recent helioseismological observations~\cite{BPBC}.

All five solar neutrino experiments~\cite{HS,GALLEX,SAGE,KK,SK}
observe a solar neutrino flux which is smaller than predicted by the
SSMs. In order to understand this discrepancy it has been suggested
that neutrinos are endowed with properties which are not present in
the standard electroweak theory~\cite{30years}.  These new properties
allow the electron neutrinos to be converted along their way from the
center of the sun to the detectors on earth into different neutrino
flavors, i.e. into muon, tau, or possibly sterile~\cite{sterile}
neutrinos.  The fact that the terrestrial experiments are less
sensitive to these neutrino flavors explains the observed lower
counting rates.  The most plausible solution is that neutrinos are
massive and there is mixing in the lepton sector.  Then neutrino
oscillations in vacuum~\cite{vacuum} or matter~\cite{wolf,MS} (the
Mikheyev-Smirnov-Wolfenstein (MSW) effect) can explain the deficit of
observed neutrinos with respect to the predictions of the
SSM~\cite{BKS,GHPV,FLMP}.

In his seminal paper Wolfenstein~\cite{wolf} observed that \NSNI\
(NSNI) with matter can also generate neutrino oscillations.  In
particular this mechanism could be relevant to solar neutrinos
interacting with the dense solar matter along their path from the core
of the sun to its surface~\cite{GMP,BPW,ER,DR,FL94,KB,Bergmann}.  
In this case the flavor changing neutrino interactions (FCNI) 
are responsible for the off-diagonal elements in the
neutrino propagation matrix (similar to the $\Delta m^2 \sin^2
2\theta$ term induced by vacuum mixing). For massless neutrinos
resonantly enhanced conversions can occur due to an interplay between
the standard electroweak neutrino interactions and non-universal 
flavor diagonal neutrino interactions (FDNI)
with matter~\cite{GMP,Valle87}.

While many extensions of the standard model allow for massive
neutrinos, it is important to stress that also many New Physics models
predict new neutrino interactions.  The minimal supersymmetric
standard model without $R$-parity has been evoked as an explicit model
that could provide the FCNI and FDNI needed for this mechanism.
Systematic studies of the data demonstrated that resonantly enhanced
oscillations induced by FCNI and FDNI for massless
neutrinos~\cite{BPW,KB}, or FCNI in combination with massive
neutrinos~\cite{BPW,Bergmann} can solve the solar neutrino problem.

In this paper we investigate the current status of the solution to the
solar neutrino problem based on NSNI, which is briefly reviewed in
section~\ref{formalism}. In the first part of our study
(section~\ref{analysis}) we present a comprehensive statistical
analysis of this solution. Our analysis comprises both the measured
total rates of Homestake~\cite{HS}, GALLEX~\cite{GALLEX},
SAGE~\cite{SAGE} and SuperKamiokande~\cite{SK} and, for the first time
in the context of NSNI, the full SuperKamiokande data set
(corresponding to 825 effective days of operation) including the
recoil electron spectrum and the day-night asymmetry.  We have not
included in our $\chi^2$ analysis the seasonal variation but we will
comment on this effect.  For the solar input we take the solar
neutrino fluxes and their uncertainties as predicted in the standard
solar model by Bahcall and Pinsonneault (hereafter BP98
SSM)~\cite{BBP}. The BP98 SSM includes helium and heavy elements
diffusion, as well as the new recommended value~\cite{Adel} for the
low energy $S$-factor, $S_{17} = 19^{+4}_{-2}$~eV~b. We also study the
dependence of the allowed parameter space on the high energy $^8$B
neutrino flux, by varying the flux normalization as a free parameter.

In the second part of our study for the first time a systematic,
model-independent investigation of the phenomenological constraints on
FCNI and new non-universal FDNI relevant for solar neutrinos is
presented (section~\ref{constraints}). Our two main goals are: a) to
find out whether NSNI can be sufficiently large to provide a viable
solution to the solar neutrino problem, and b) to study various kinds
of new interactions in order to single out those New Physics models
that can provide such interactions. Since the typical energy scales
relevant for solar neutrinos are lower than the weak interaction scale
and therefore lower than any New Physics scale, it is sufficient to
discuss the effective operators induced by heavy boson exchange that
allow for non-standard neutrino scattering off quarks or electrons.
These operators are related by the $SU(2)_L$ symmetry of the standard
electroweak theory to operators that induce anomalous contributions to
leptonic decays. Since $SU(2)_L$ violation cannot be large for New
Physics at or above the weak scale, one can use the upper bounds on
lepton flavor violating decays or on lepton universality violation to
put model-independent bounds on the relevant non-standard neutrino
interactions.

We find that \NSNI\ can provide a good fit to the solar neutrino data
if there are rather large non-universal FDNI (of order $0.5 \, G_F$)
and small FCNI (of order a few times $10^{-3} \, G_F$).  Our
phenomenological analysis indicates that FCNI could only be large
enough to provide $\nu_e \to \nu_\tau$ transitions, while $\nu_e \to
\nu_\mu$ transitions are not relevant for the solution of the solar
neutrino problem, because of strong experimental constraints.  Large
FDNI can only be induced by an intermediate doublet of $SU(2)_L$ (a
scalar or a vector boson) or by a neutral vector singlet.  We conclude
that the minimal supersymmetric model with broken
$R$-parity~\cite{SUSYwoR} is the favorite model for this scenario.

In section~\ref{futureexperiments} we discuss how to to confirm or
exclude the solution to the solar neutrino problem based on \NSNI\ by
future experiments.  We argue that the magnitudes of FCNI parameters
necessary for $\nu_e \to \nu_\tau$ conversion in the sun could be
tested independently by the upcoming $B$-factories.  Finally, we
discuss briefly the possibility of distinguishing this solution from
the others by future solar and long-baseline neutrino oscillation
experiments.

\section{Neutrino flavor conversion induced by non-standard 
         neutrino interactions}
\label{formalism}

Any model beyond the standard electroweak theory that gives rise to
the processes
\beqs
\nu_e \, f      &\to& \nu_\ell \, f   \,, \label{FCNI} \\
\nu_\alpha \, f &\to& \nu_\alpha \, f \,, \label{FDNI}
\eeqs
where (here and below) $f=u,d,e$ and $\ell=\mu, \tau$ and $\alpha=e,
\mu, \tau$, is potentially relevant for neutrino oscillations in the
sun, since these processes modify the effective mass of neutrinos
propagating in dense matter.

The evolution equations for massless neutrinos that interact with
matter via the standard weak interactions and the non-standard
interactions in~(\ref{FCNI}) and~(\ref{FDNI}) is given
by~\cite{GMP,BPW}:
\beq \label{eq-of-motion} 
i\frac{d}{dr} \pmatrix{A_e(r) \cr A_\ell(r)} = 
\sqrt2 G_F
\pmatrix{n_e(r)                       & \epsilon_{\nu_\ell}^f n_f(r)  \cr
         \epsilon_{\nu_\ell}^f n_f(r) & {\epsilon'}_{\nu_\ell}^f n_f(r) \cr} 
\, \pmatrix{A_e(r) \cr A_\ell(r)} \,,
\eeq
where $A_e(r)$ and $A_\ell(r)$ are, respectively, the probability
amplitudes to detect a $\nu_e$ and $\nu_\ell$ at position $r$.  For
neutrinos that have been coherently produced as $\nu_e$ in the solar
core at position $r_0$, the equations in~(\ref{eq-of-motion}) are
subject to the boundary conditions $A_e(r_0)=1$ and $A_\ell(r_0)=0$.
While $W$-exchange of $\nu_e$ with the background electrons gives rise
to the well known forward scattering amplitude $\sqrt2 G_F n_e(r)$,
the FCNI in~(\ref{FCNI}) induce a flavor changing forward scattering
amplitude $\sqrt2 G_F \epsilon_{\nu_\ell}^f n_f(r)$ and the
non-universal FDNI are responsible for the flavor diagonal entry
$\sqrt2 G_F {\epsilon'}_{\nu_\ell}^f n_f(r)$ in
eq.~(\ref{eq-of-motion}).  Here
\beq
n_f(r)=\left\{\matrix{n_n(r) + 2n_p(r) & f=u \cr
                      2n_n(r) + n_p(r) & f=d} \right. 
\eeq 
is the respective fermion number density at position $r$ in terms of
the proton [neutron] number density $n_p(r)$ [$n_n(r)$] and
\beq \label{defeps}
\varepsilon  = \epsilon_{\nu_\ell}^f \equiv 
 {G_{\nu_e \nu_\ell}^f \over G_F} \andthis
\varepsilon' = {\epsilon'}_{\nu_\ell}^f \equiv 
 {G_{\nu_\ell \nu_\ell}^f - G_{\nu_e \nu_e}^f \over G_F} \,,
\eeq
describe, respectively, the relative strength of the FCNI
in~(\ref{FCNI}), and the new flavor diagonal, but non-universal
interactions in~(\ref{FDNI}). $G_{\nu_\alpha \nu_\beta}^f$ ($\alpha,
\beta = e, \mu, \tau$) denotes the effective coupling of the
four-fermion operator
\beq \label{Onu}
{\cal O}_\nu^f \equiv (\lbar{\nu_\alpha} \, \nu_\beta) \, (\bar f \, f) \,.
\eeq
that gives rise to such interactions. The Lorentz structure of ${\cal
  O}_\nu^f$ depends on the New Physics that induces this operator.
Operators which involve only left-handed neutrinos (and which conserve
total lepton number $L$) can be decomposed into a $(V-A) \otimes
(V-A)$ and a $(V-A) \otimes (V+A)$ component.  (Any single New Physics
contribution that is induced by chiral interactions yields only one of
these two components.) It is, however, important to note that only the
vector part of the background fermion current affects the neutrino
propagation for an unpolarized medium at rest~\cite{wolf,BGN}.  Hence
only the $(V-A) \otimes (V)$ part of ${\cal O}_\nu^f$ is relevant for
neutrino oscillations in normal matter.  One mechanism to induce such
operators is due to the exchange of heavy bosons that appear in
various extensions of the standard model.  An alternative mechanism
arises when extending the fermionic sector of the standard model and
is due to $Z$-induced flavor-changing neutral currents (FCNCs). For a
discussion of $Z$-induced FCNC effects on solar neutrinos, see
Refs.~\cite{Z-FCNC,BergmannKagan}.

A resonance occurs when the diagonal entries of the evolution matrix
in eq.~(\ref{eq-of-motion}) coincide at some point $r_{res}$ along the
trajectory of the neutrino, leading to the resonance condition
\beq 
{\epsilon'}_{\nu_\ell}^f n_f(r_{res}) = n_e(r_{res}) \,.  
\label{res}
\eeq
An immediate consequence is that new FDNI for $f=e$ alone cannot
induce resonant neutrino flavor conversions.

As we will see in section~\ref{constraints} only $\nu_e \to \nu_\tau$
conversions are compatible with the existing phenomenological
constraints on $\epsilon_{\nu_\ell}^f$ and ${\epsilon'}_{\nu_\ell}^f$.
We note that in the minimal supersymmetric standard model with broken
$R$-parity~\cite{SUSYwoR} the relevant parameters are given by
\beq
\epsilon_{\nu_\tau}^d    = {{\lambda'_{331}}^* \cdot \lambda'_{131} \over 
                              4 M_{\tilde b}^2 \sqrt2 \, G_F} \,, \andthis
{\epsilon'}_{\nu_\tau}^d = {|\lambda'_{331}|^2 - |\lambda'_{131}|^2 \over
                              4 M_{\tilde b}^2 \sqrt2 \, G_F} \,,
\eeq
in terms of the trilinear couplings $\lambda'_{ijk}$ and the bottom
squark mass $M_{\tilde b}$.

The neutrino evolution matrix in eq.~(\ref{eq-of-motion}) vanishes in
vacuum and is negligibly small for the matter densities of the earth's
atmosphere.  Therefore the probability of finding an electron neutrino
arriving at the detector during day time is easily obtained by
evolving the equations in~(\ref{eq-of-motion}) from the neutrino
production point to the solar surface.  Furthermore, typically there
are many oscillations between the neutrino production and detection
point and a resonance. Therefore the phase information before and
after the resonance is usually lost after integration over the
production and detection region and one may use classical survival
probabilities. Then at day time we have~\cite{BPW}
\beq \label{day}
P_{\nu_e \to \nu_e}^{\rm day} = |A_e(r_s)|^2 \simeq 
\frac{1}{2} + (\frac{1}{2} - P_c) \cos 2\theta_m^p ~\cos 2\theta_m^s \,,
\eeq
where $r_s$ is the solar surface position and in the analytic
expression in eq.~(\ref{day}) we denote by $\theta_m^p$ and
$\theta_m^s$, respectively, the effective, matter-induced mixing at
the neutrino production point and at the solar surface. In terms of
the New Physics parameters $\varepsilon$, $\varepsilon'$ and the
fermion densities the effective mixing is given by~\cite{GMP,BPW}
\beq
\tan 2\theta_m = {2\epsilon_{\nu_\ell}^f n_f \over 
                  {\epsilon'}_{\nu_\ell}^f n_f - n_e} \,.
\eeq
Note that $\tan 2\theta_m = 2\epsilon_{\nu_\ell}^e /
({\epsilon'}_{\nu_\ell}^e-1)$ is constant for $f=e$. $P_c$ is the
level crossing probability. The approximate Landau-Zener expression
is~\cite{GMP,BPW}
\beq \label{Pc}
P_c = 
\exp \left[-\pi \gamma / 2 \right] ~~~~~{\rm with}~~~~~
\gamma = 4 \sqrt2 G_F 
\left| {(\epsilon_{\nu_\ell}^f/{\epsilon'}_{\nu_\ell}^f)^2 
 \over {\epsilon'}_{\nu_\ell}^f} \cdot
{n_e \over {d \over dx} \left(n_f \over n_e \right)} \right|_{res} \,.
\eeq
When neutrinos arrive at the detector during the night, a modification
of the survival probability has to be introduced since the
non-standard neutrino interactions with the terrestrial matter may
regenerate electron neutrinos that have been transformed in the sun.
Assuming that the neutrinos reach the Earth as an incoherent mixture
of the effective mass-eigenstates $\nu_1$ and $\nu_2$ the survival
probability during night-time can be written as~\cite{eartheffect}:
\beq \label{night}
P_{\nu_e \to \nu_e}^{\rm night} = \frac{P_{\nu_e \to \nu_e}^{\rm day} - 
\sin^2\theta_m^s+P_{2e}(1-2P_{\nu_e \to \nu_e}^{\rm day})}{\cos 2\theta_m^s} 
\,.
\eeq
Here $P_{2e}$ is the probability of a transition from the state $\nu_2$ 
to the flavor eigenstate $\nu_e$ along the neutrino path in the Earth. 

For our analysis we assume a step function profile for the
Earth matter density, which has been shown to be a good approximation
in other contexts (see e.g. Ref.~\cite{Freund} for a recent analysis
of matter effects for atmospheric neutrinos). 
Then the earth matter
effects on the neutrino propagation correspond to a parametric
resonance and can be calculated analytically~\cite{Akhmedov},
\beq \label{pe2earth}
P_{2e}=\sin^2\theta_m^s + W_1^2\cos 2\theta_m^s + 
 W_1 W_3 \sin 2\theta_m^s \,, 
\eeq
where the parameters $W_1$ and $W_3$ contain all the information of
the Earth density and are defined in Ref.~\cite{Akhmedov}.  (The only
difference is that in our case also the off-diagonal element of the
neutrino evolution matrix varies when the neutrino propagates through
the earth matter.)

It is this interaction with the terrestrial matter that can produce a
day-night variation of the solar neutrino flux and, consequently, a
seasonal modulation of the data. (Note that this seasonal variation is
of a different nature than the one expected for vacuum oscillations
from the change of the baseline due to the eccentricity of the
earth's orbit around the sun.)

\section{Analysis of the solar neutrino data}
\label{analysis}

In this section we present our analysis of the solution to the solar
neutrino problem based on neutrino flavor conversions induced by NSNI
in matter.  Our main goal is to determine the values of $\varepsilon$
and $\varepsilon'$ that can explain the experimental observations
without modifying the standard solar model predictions.

\subsection{Rates}

First we consider the data on the total event rate measured by the
Chlorine (Cl) experiment~\cite{HS}, the Gallium (Ga) detectors
GALLEX~\cite{GALLEX} and SAGE~\cite{SAGE} and the water Cherenkov
experiment SuperKamiokande (SK)~\cite{SK}. We compute the allowed
regions in parameter space according to the BP98 SSM~\cite{BBP} and
compare the results with the regions obtained for an arbitrary
normalization $f_B$ of the high energy neutrino $^8$B neutrino fluxes.

We use the minimal $\chi^2$ statistical treatment of the data
following the analyses of Refs.~\cite{FLM,FL}. Our $\chi^2$ function
is defined as follows:
\beq \label{chi}
\chi^2_R(\varepsilon,\varepsilon' [,f_B]) = \sum_{i,j=1,...,4}
 \left(R_i^{th}(\varepsilon,\varepsilon' [,f_B])-R_i^{obs} \right) \, 
 \left[\sigma^2_R \right]^{-1}_{ij} \,
 \left(R_j^{th}(\varepsilon,\varepsilon' [,f_B])-R_j^{obs} \right) \,,
\eeq
where $R_i^{th}$ and $R_i^{obs}$ denote, respectively, the predicted
and the measured value for the event rates of the four solar
experiments ($i$ = Cl, GALLEX, SAGE, SK). The error matrix $\sigma_R$
contains both the experimental (systematic and statistical) and the
theoretical errors.

\begin{figure}[ht]
\vspace{-1.5cm}
\vglue 1.5cm \hglue -1.5cm
\epsfig{file=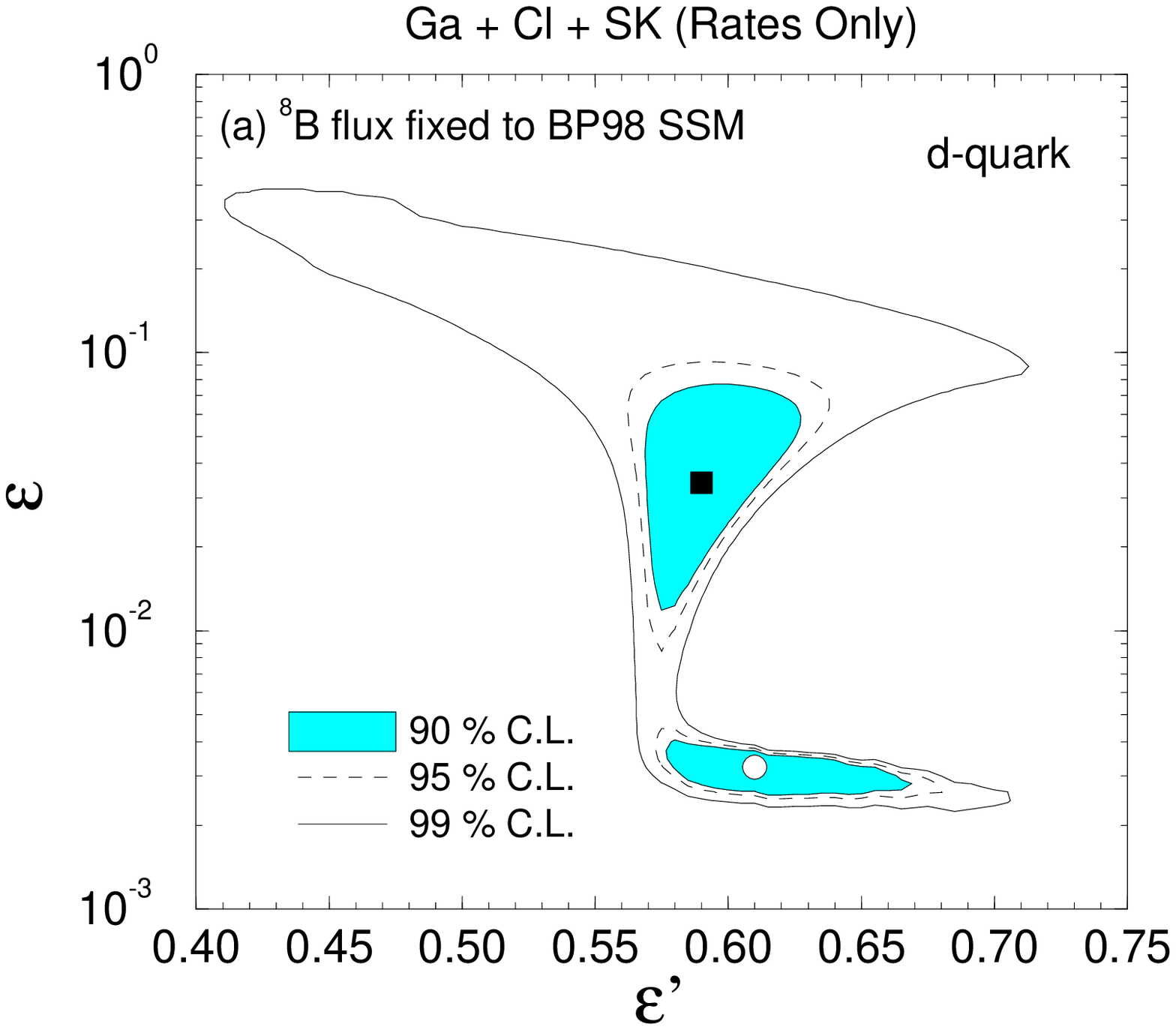,width=9.5cm}
\vglue -7.75cm \hglue 7.0 cm 
\epsfig{file=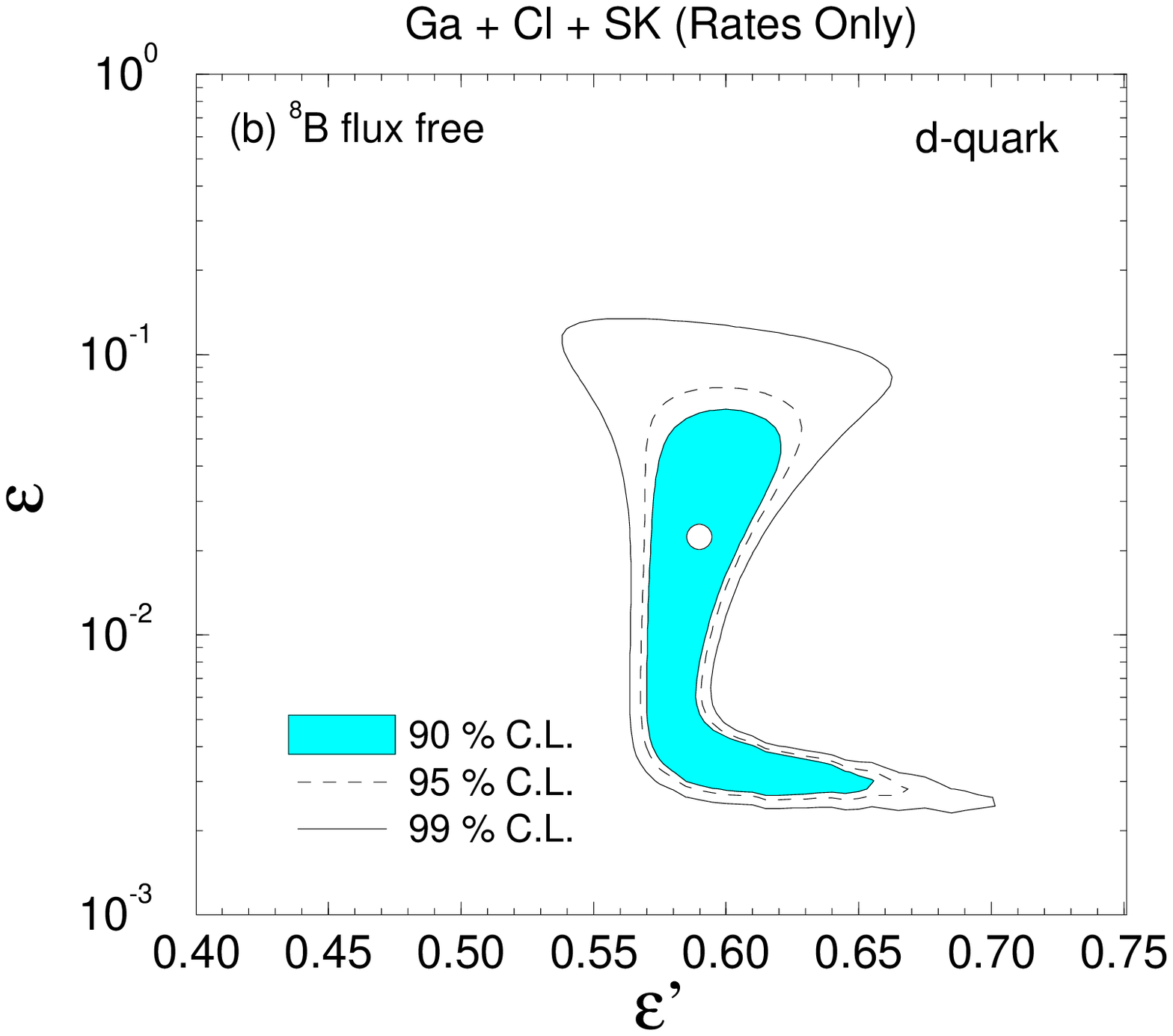,width=9.5cm}
\caption{\noindent  
  Region of $\varepsilon=\epsilon_\nu^d$ and $\varepsilon' =
  {\epsilon'}_\nu^d$ which can explain the total rates measured by the
  Homestake, GALLEX, SAGE and SuperKamiokande solar neutrino
  experiments in terms of non-standard neutrino interactions with
  $d$-quarks.  (a) The best fit (indicated by the open circle) is
  obtained for $(\varepsilon, \varepsilon') = (0.0032, 0.610$) with
  $\chi^2_{min} = 2.44$ for $4-2=2$ \DOF.  A second (local) $\chi^2$
  minimum (indicated by the solid square) is found at $(\varepsilon,
  \varepsilon') = (0.034, 0.610)$ with $\chi^2 = 2.63$. (b) Allowing
  for an arbitrary $^8$B flux normalization $f_B$, the best fit
  (indicated by the open circle) is obtained for $(\varepsilon,
  \varepsilon') = (0.022, 0.590)$ and $f_B = 1.36$ with $\chi^2_{min}
  = 0.91$ for $4-3=1$ \DOF.}
\label{rates-d}
\end{figure}

In Fig.~\ref{rates-d} the allowed regions in the parameter space of
$\epsilon_\nu^d$ and ${\epsilon'}_\nu^d$ for neutrino scattering off
$d$-quarks are shown at 90, 95 and 99\,\% confidence level~(\CL). In
Fig.~\ref{rates-d}a, the $^8$B flux is fixed by the BP98 SSM
prediction ($f_B=1$). The best fit point of this analysis is found at
\beq
\epsilon_\nu^d    = 3.2 \times 10^{-3} \andthis
{\epsilon'}_\nu^d = 0.61 \,,
\eeq
with $\chi^2_{min} = 2.44$ for $4-2=2$ degrees of freedom (\DOF).
Allowing an arbitrary $^8$B flux normalization, a different best fit
point is obtained for $(\epsilon_\nu^d, {\epsilon'}_\nu^d) = (2.2
\times 10^{-2}, 0.59)$ and $f_B=1.36$ with $\chi^2_{min} = 0.91$ for
$4-3=1$ \DOF. The result of this analysis is shown in
Fig.~\ref{rates-d}b. (Effects due to deviations of the hep neutrino
flux from the standard solar model prediction are expected to be
less significant and we do not consider them in this work.)

\begin{figure}[ht]
\vspace{-1.5cm}
\vglue 1.5cm \hglue -1.5cm
\epsfig{file=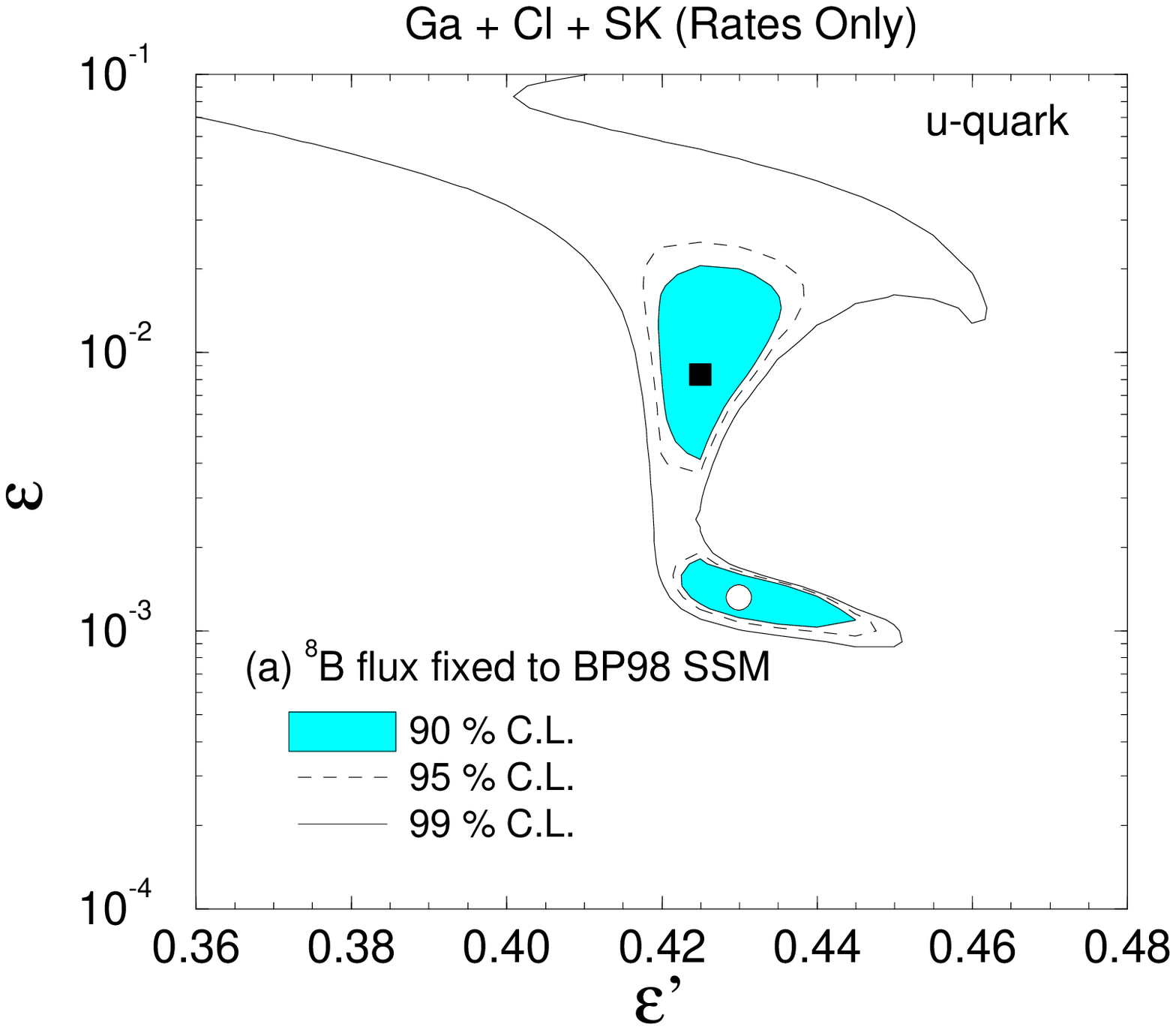,width=9.5cm}
\vglue -7.75cm \hglue 7.0 cm 
\epsfig{file=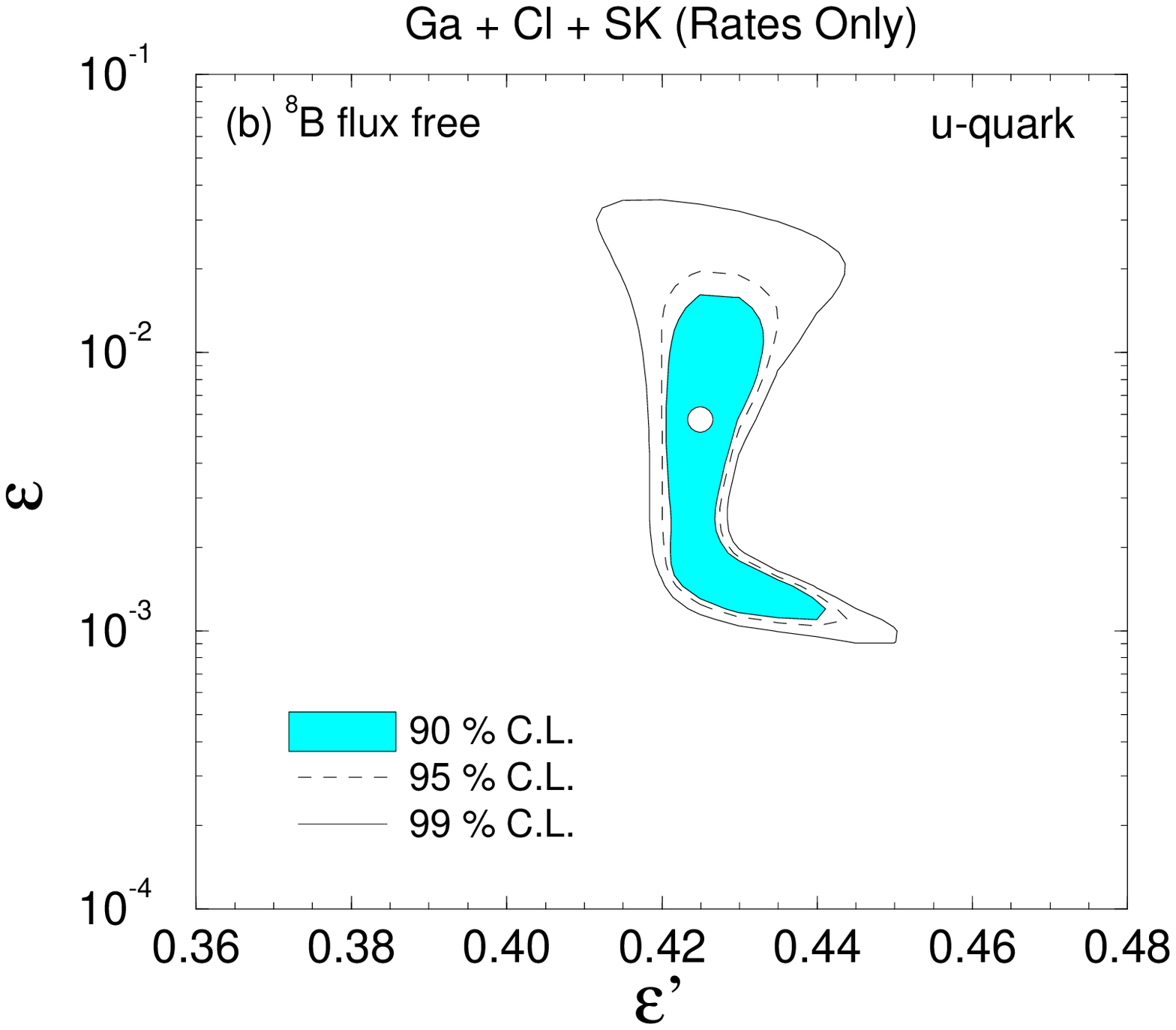,width=9.5cm}
\caption{\noindent  
  Same as in Fig.~\ref{rates-d} but for $u$-quarks.  (a) The best fit
  (indicated by the open circle) is obtained for $(\varepsilon,
  \varepsilon') = (0.0013, 0.430$) with $\chi^2_{min} = 2.75$ for
  $4-2=2$ \DOF.  A second (local) $\chi^2$ minimum (indicated by the
  solid square) is found at $(\varepsilon, \varepsilon') = (0.0083,
  0.425)$ with $\chi^2 = 2.70$. (b) Allowing for an arbitrary $^8$B
  flux normalization $f_B$, the best fit (indicated by the open
  circle) is obtained for $(\varepsilon, \varepsilon') = (0.0058,
  0.425)$ and $f_B = 1.34$ with $\chi^2_{min} = 0.96$ for $4-3=1$
  \DOF.}  
\vglue -1.5cm
\label{rates-u}
\end{figure}

In Fig.~\ref{rates-u} the allowed regions in the parameter space of
$\epsilon_\nu^u$ and ${\epsilon'}_\nu^u$ for neutrino scattering off
$u$-quarks are shown at 90, 95 and 99\,\%~\CL. In Fig.~\ref{rates-u}a,
the $^8$B flux is fixed by the BP98 SSM prediction.  The best fit
point of this analysis is found at
\beq
\epsilon_\nu^u    = 1.32 \times 10^{-3} \andthis
{\epsilon'}_\nu^u = 0.43 \,,
\label{best_rates_u}
\eeq
with $\chi^2_{min} = 2.64$ for two \DOF.  Allowing an arbitrary $^8$B
flux normalization $f_B$, a different best fit point is obtained for
$(\epsilon_\nu^u, {\epsilon'}_\nu^u) = (5.8 \times 10^{-3}, 0.425)$
and $f_B=1.34$ with $\chi^2_{min} = 0.96$ for one \DOF. The result of
this analysis is shown in Fig.~\ref{rates-u}b.

\begin{figure}[ht]
\vglue 0.5cm
\centerline{\epsfig{file=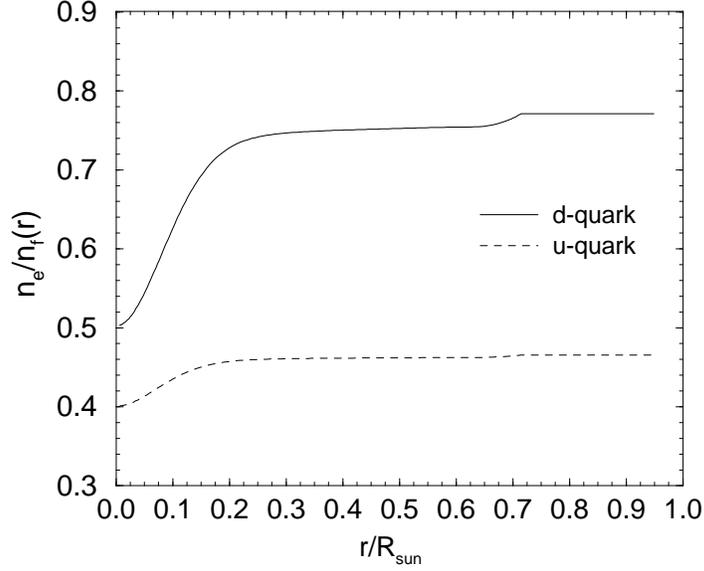,width=10.0cm}}
\caption{\noindent  
  The ratio of the number density of electron to that of $d$ and
  $u$-quarks in the sun $n_e(r)/n_f(r)$ ($f = d,u$), is plotted as a
  function of the distance from the solar center.}
\label{ne_nf_ratio}
\end{figure}

It is remarkable that the neutrino flavor conversion mechanism based
on NSNI provides quite a good fit to the total rates despite the fact
that the conversion probabilities~(\ref{day}) and~(\ref{night}) do not
depend on the neutrino energy. This is unlike the case of the vacuum
and the MSW conversion mechanisms which provide the appropriate energy
dependence to yield a good fit. For NSNI the only way to distinguish
between neutrinos of different energies is via the position of the
resonance $r_{res}$. Note that according to eq.~(\ref{res}), $r_{res}$
is a function of $\varepsilon'$ only.  As can be seen in
Fig.~\ref{ne_nf_ratio}, $n_e / n_f$ ($f = d,u$) is a smooth and
monotonic function of the distance from the solar center $r$, allowing
to uniquely determine $r_{res}$ for a given value of $\varepsilon'$.
From Fig.~\ref{ne_nf_ratio} it follows that a resonance can only occur
if $\epsilon'_d \in [0.50, 0.77]$ for NSNI with $d$-quarks or
$\epsilon'_u \in [0.40, 0.46]$ for NSNI with $u$-quarks.  For both
cases the major part of these intervals corresponds to $r_{res} \lsim
0.2 \, R_\odot$ ($R_\odot$ being the solar radius). For
$\epsilon'_{d,u}$ within the 90\,\%~\CL\ regions (indicated in
Fig.~\ref{rates-d} and Fig.~\ref{rates-u}) we find $r_{res} \approx
0.1 ~ R_\odot$.  Since the nuclear reactions that produce neutrinos
with higher energies in general take place closer to the solar center
(see chapter 6 of Ref.~\cite{JNB} for the various spatial
distributions of the neutrino production reactions), a resonance
position close to the solar center implies that predominantly the high
energy neutrinos are converted by a resonant transition. For $r_{res}
\approx 0.1 ~ R_\odot$ practically all $^8$B-neutrinos cross the
resonance layer, fewer $^7$Be-neutrinos pass through the resonance,
while most of the $pp$-neutrinos are not be affected by the resonance
since their production region extends well beyond the resonance layer.
Therefore for most of the allowed region in Fig.~\ref{rates-d} and
Fig.~\ref{rates-u} we have
\beq \label{P_relation}
\langle P(^8\text{B}) \rangle < 
\langle P(^7\text{Be}) \rangle < 
\langle P(pp) \rangle \,. 
\eeq
We note that the above relation is still valid when taking into
account that a significant fraction of the $pp$ neutrinos crosses the
resonance layer twice, if they are produced just outside resonance.
This is -- roughly speaking -- because a $\nu_e$ which undergoes a
resonant flavor transition when entering the solar interior at
$r_{res}$ is reconverted into a $\nu_e$ at the second resonance when
it emerges again from the solar core. In our numerical calculations we
properly take into account the effects of such double resonances.

An immediate consequence of the relation in eq.~(\ref{P_relation}) is
that as long as $f_B=1$ the NSNI solution predicts that $R_{SK} <
R_{Cl} < R_{Ga}$, which is inconsistent with the observed hierarchy of
the rates, $R_{Cl} < R_{SK} < R_{Ga}$, leading to a somewhat worse fit
than the standard MSW solutions. However when treating $f_B$ as a free
parameter, for $f_B \sim 1.3-1.4$ the SK rate is sufficiently enhanced
to give the correct relation between the rates. In this case also the
neutral current contribution from $\nu_{\mu,\tau} \, e^-$ scattering
is increased due to a larger $\nu_{\mu,\tau}$ flux, which is
consistent with the Super-Kamiokande observations. We find that for
the best fit points for $(\varepsilon,\varepsilon')$ in
Figs.~\ref{rates-d}b and~\ref{rates-u}b and $f_B \sim 1.35$ the
survival probability for $^8$B, $^7$Be and $pp$-neutrinos are $\sim
0.24$, $0.4$ and $0.7$, respectively.

\begin{figure}[ht]
\vglue -0.5cm
\vglue 1.5cm \hglue -1.6cm
\epsfig{file=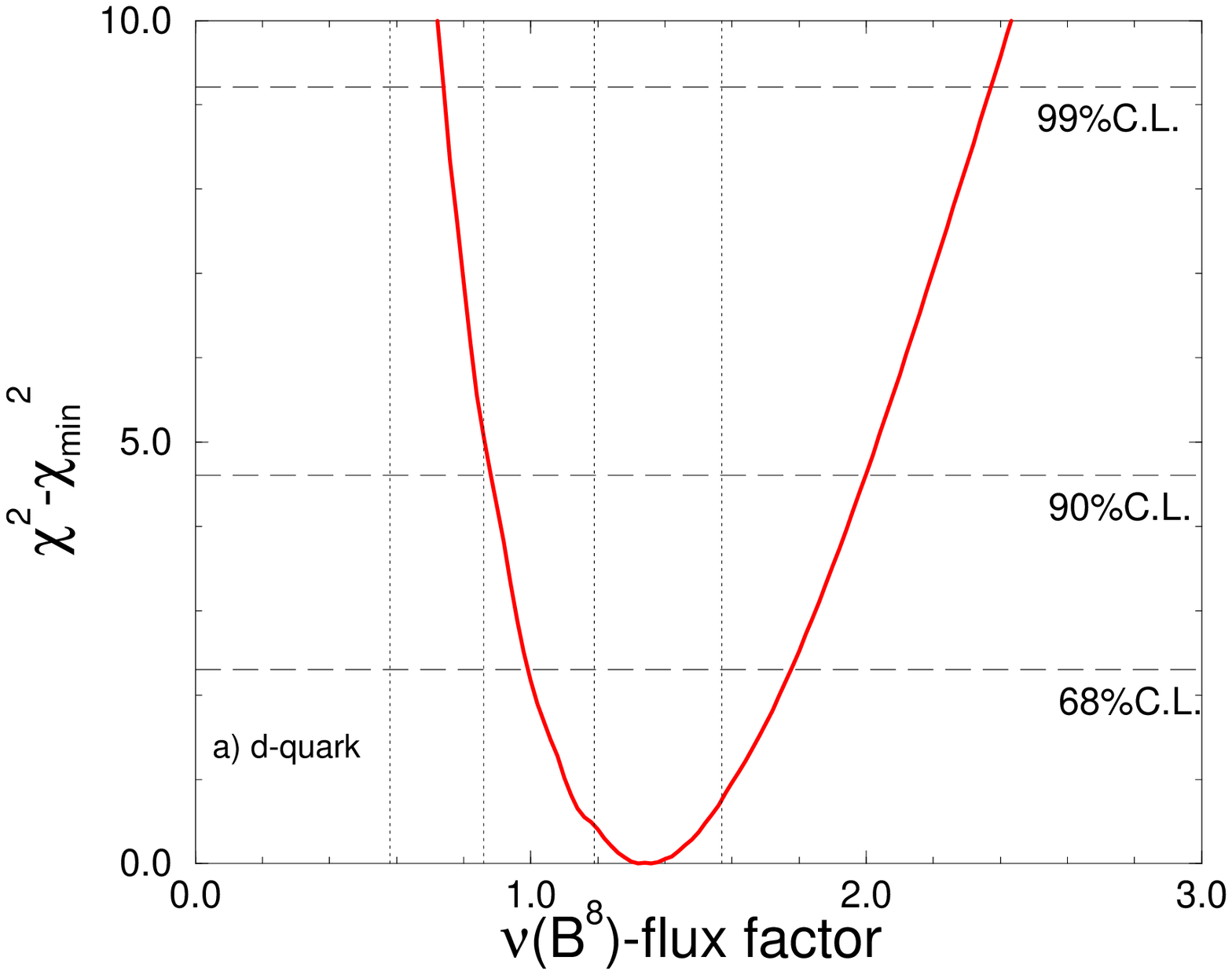,width=9.5cm}
\vglue -7.8cm \hglue 7.1 cm 
\epsfig{file=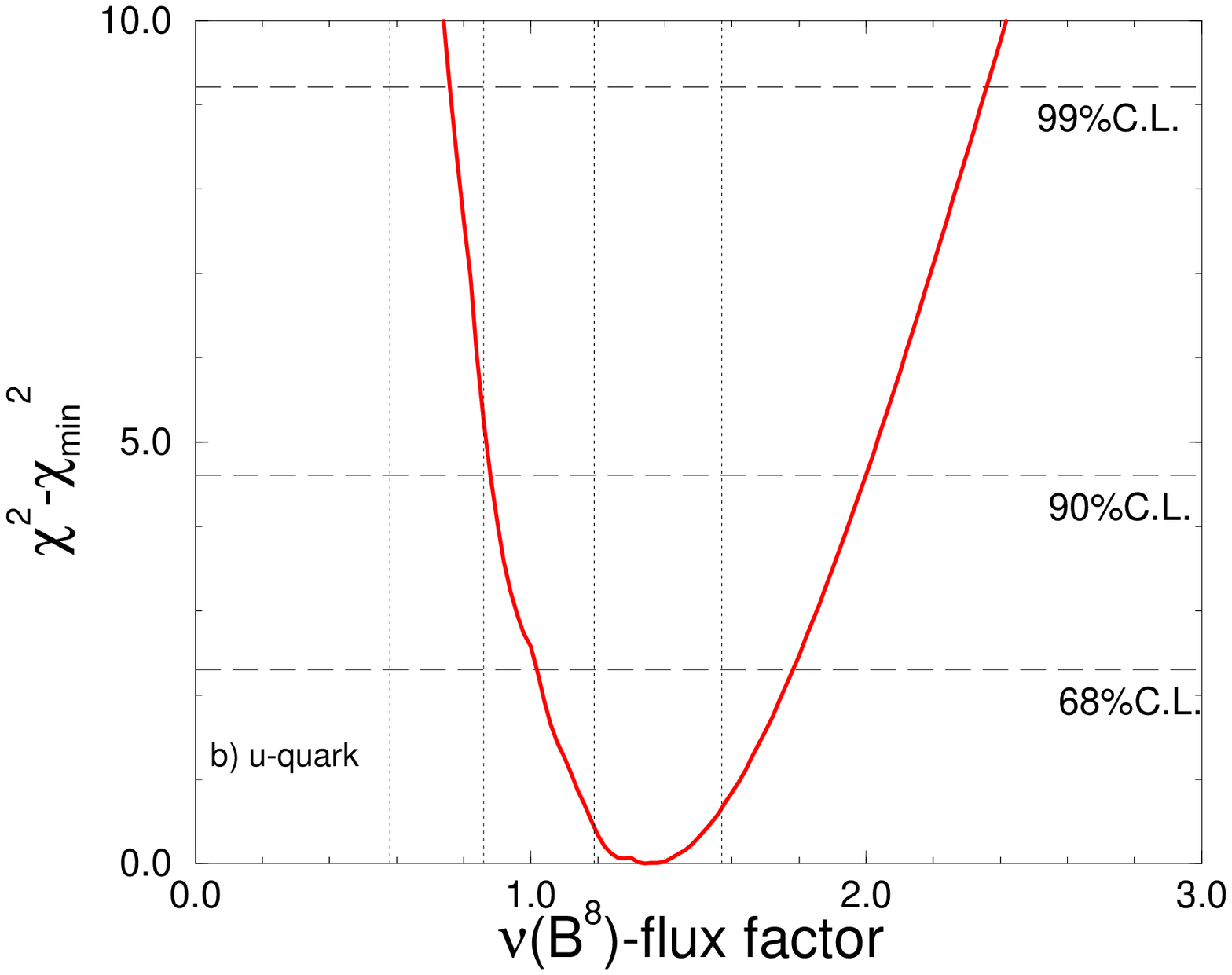,width=9.5cm}
\caption{\noindent  Minimum $\chi^2$ calculated as a function of 
  the boron neutrino flux for (a) d-quark and (b) u-quark.} 
\vglue 0.5cm
\label{factor}
\end{figure}

In Fig.~\ref{rates-d}b and Fig.~\ref{rates-u}b the $^8$B neutrino flux
normalization $f_B$ has been varied as a free parameter in order to
study the dependence of the allowed parameter space on the high energy
neutrino flux. It is interesting to note that the allowed regions in
these figures do not completely contain those in Fig.~\ref{rates-d}a
and Fig.~\ref{rates-u}a where the boron flux and its uncertainty is
determined by the BP98 SSM. In order to explain this apparently
inconsistent result we have have plotted $\chi^2$ as a function of
$f_B$ in Fig.~\ref{factor} allowing $f_B$ to vary within a
sufficiently broad interval ($0 < f_B < 100$) for every point in the
$(\varepsilon, \varepsilon')$ parameter space.  The horizontal lines
indicate the 68, 90 and 99\,\%~\CL\ limits for two \DOF. The
intersection of these lines with the $\chi^2$ curve determine the
relevant ranges of the boron flux allowed by the experimental data.
The vertical dotted lines indicate the $1\sigma$ and $3\sigma$ ranges
of the boron neutrino flux in the BP98 SSM.

Note that the $\chi^2$ minima are obtained for a boron flux
significantly larger ($f_B \sim 1.35$) than the one predicted by the
SSM ($f_B = 1.0$), as we already anticipated in the discussion of
eq.~(\ref{P_relation}).  Moreover, from Fig.~\ref{factor} it follows
that for $f_B < 1$ the fit to the experimental data imposes stronger
constraints on the boron flux than the SSM.  For $f_B > 1$ the
situation is exactly the opposite.  Therefore the effect of relaxing
$f_B$ from its SSM value is that regions in the
$(\varepsilon,\varepsilon')$ parameter space where the averaged
survival probability for $^8$B neutrinos, $\langle P(^8\text{B})
\rangle$ is smaller can be easily compensated by a larger boron flux
and obtain a lower value for $\chi^2_R$.  On the other hand regions
where $\langle P(^8\text{B}) \rangle$ is rather large require a small
boron flux which is more difficult to achieve when eliminating the SSM
constraint on $f_B$.  This is the main mechanism behind the changes of
the allowed regions upon relaxing the SSM constraint on $f_B$. It
explains why the regions with large $\varepsilon$ are allowed in
Fig.~\ref{rates-d}a and Fig.~\ref{rates-u}a and are ruled out in
Fig.~\ref{rates-d}b and Fig.~\ref{rates-u}b. Here $\langle
P(^8\text{B}) \rangle$ is rather large and a small boron flux $f_B
\sim 1$ like in the SSM is preferred to explain the data.  The
opposite occurs in the area between the two disconnected regions in
Fig.~\ref{rates-d}a and Fig.~\ref{rates-u}a.  Here $\langle
P(^8\text{B}) \rangle$ is comparatively small and therefore a larger
boron flux increases $\chi^2_R$ in this region leading to the merging
of the separated contours in Fig.~\ref{rates-d}b and
Fig.~\ref{rates-u}b when $f_B$ is treated as a free parameter.

\subsection{Zenith Angle Data}

Next, we consider the zenith angle dependence of the solar neutrino
data of the SuperKamiokande experiment. As mentioned above, NSNI with
matter may affect the neutrino propagation through the earth resulting
in a difference between the event rates during day and night time. The
data obtained by the SuperKamiokande collaboration are divided into
five bins containing the events observed at night and one bin for the
events collected during the day~\cite{Suzuki} and have been averaged
over the period of SuperKamiokande operation: 403.2 effective days for
the day events and 421.5 effective days for the night events. The
experimental results suggest an asymmetry between the total data
collected during the day ($D$) and the total data observed during the
night ($N$)~\cite{Suzuki}:
\beq 
A = 2 \, {N-D \over N+D} = 0.065 \pm 0.031 (stat.) \pm 0.013 (syst.)
\,.
\label{day/night}
\eeq

In order to take into account the earth matter effect we define the
following $\chi^2$-function that characterizes the deviations of the
six measured ($Z^{obs}_i$) from the predicted ($Z^{th}_i$) values of
the rate as a function of zenith angle:

\beq
\chi^2_Z(\varepsilon,\varepsilon',\alpha_Z) = \sum_{i=1,...,6} 
{\left(\alpha_Z Z^{th}_i(\varepsilon,\varepsilon') - Z^{obs}_i \right)^2
 \over \sigma_{Z,i}^2} \,.
\eeq
Here $\sigma_{Z,i}$ refers to the total error associated with each
zenith angle bin and we have neglected possible correlations between
the systematic errors of these bins.  Since we are only interested in
the shape of the zenith angle distribution, we have introduced an
overall normalization factor, $\alpha_Z$, which is treated as a free
parameter and determined from the fit. (Using this procedure also
prevents over-counting the data on the total event rate when combining
all available data in section~\ref{combined}).  Note that the
experimental value of the day-night asymmetry in eq.~(\ref{day/night})
is not used in the fit, since the six zenith angle bins already
include consistently all the available information about the earth
effects.

\begin{figure}[ht]
\vspace{-1.5cm}
\vglue 1.5cm \hglue -1.5cm
\epsfig{file=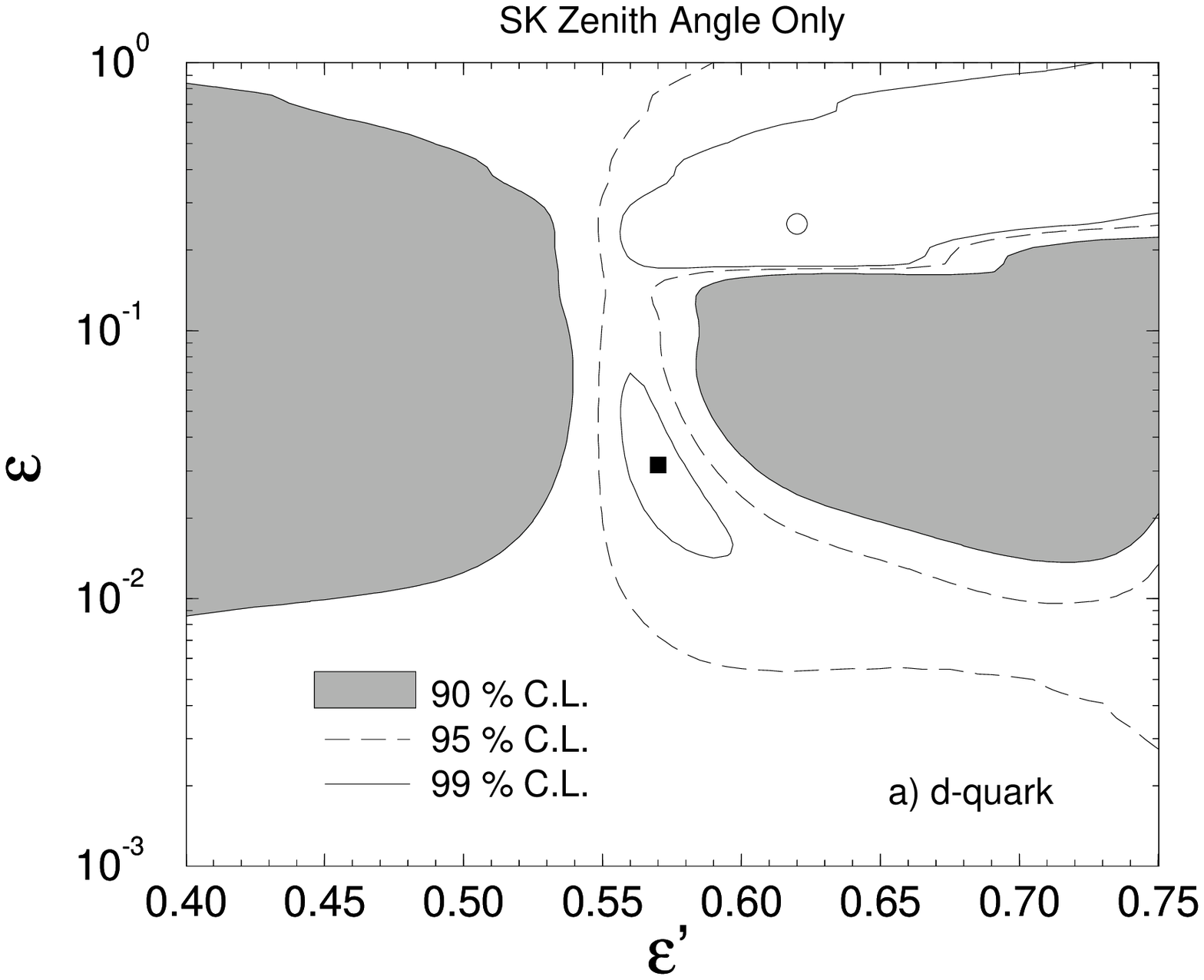,width=9.5cm}
\vglue -7.9cm \hglue 7.0 cm 
\epsfig{file=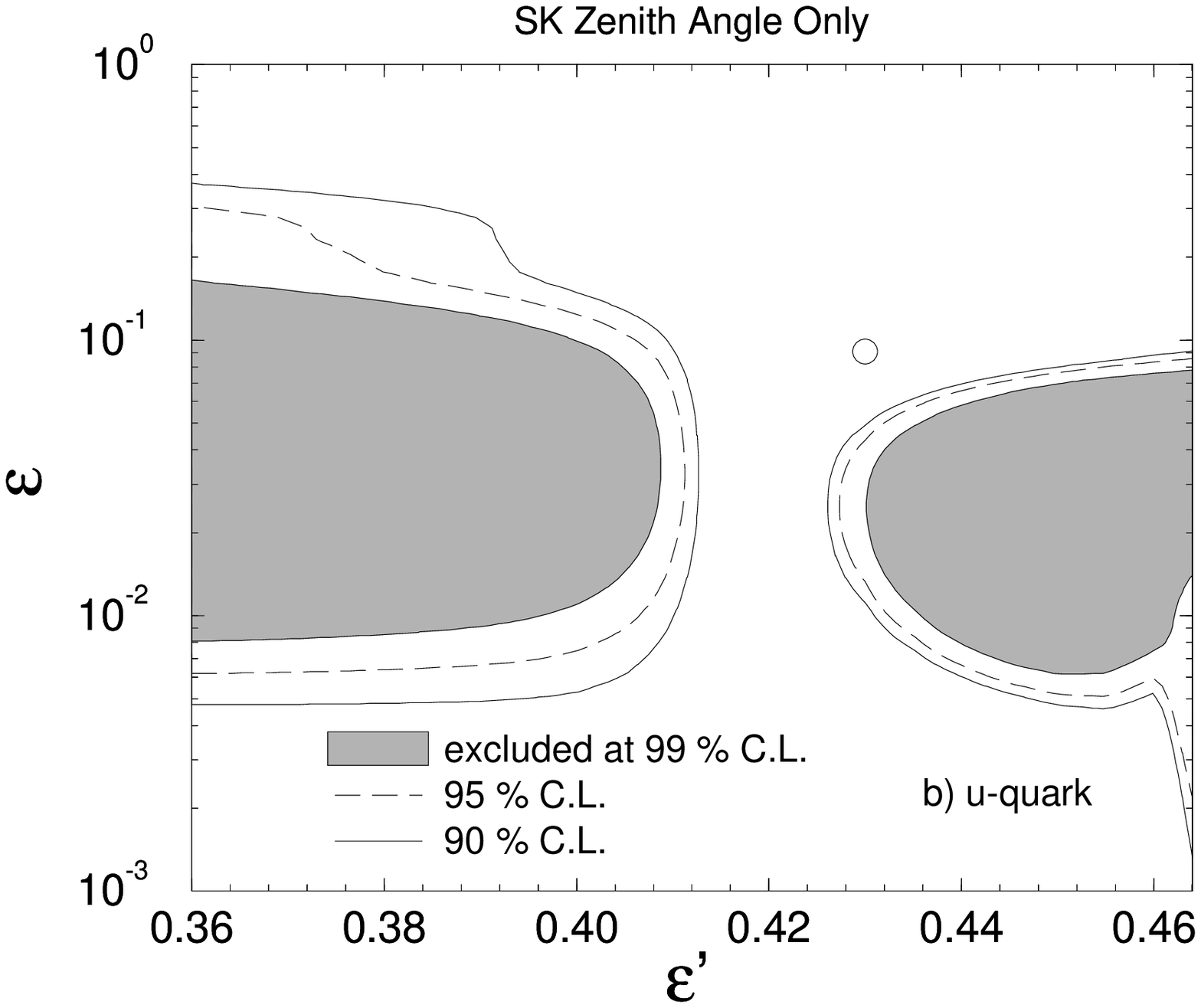,width=9.5cm}
\caption{\noindent
  Region of $\varepsilon=\epsilon_\nu^d$ and $\varepsilon' =
  {\epsilon'}_\nu^d$ which is excluded by day and night data
  (contained in $1+5$ bins) as measured by the SuperKamiokande solar
  neutrino experiment in terms of non-standard neutrino interactions
  with (a) $d$-quarks and (b) $u$-quarks.  For $d$-quarks, the best
  fit (indicated by the open circle) is obtained for $(\varepsilon,
  \varepsilon') = (0.251, 0.620)$ and $\alpha_Z = 0.819$ with $\chi^2_{min}
  = 1.10$ for $6-3=3$ \DOF.  A second (local) $\chi^2$ minimum
  (indicated by the solid squared) is found at $(\varepsilon,
  \varepsilon') = (0.0316,0.570$) and $\alpha_Z = 1.02$ with $\chi^2 =
  5.20$.  For $u$-quarks, the best fit (indicated by the open circle)
  is obtained for $(\varepsilon, \varepsilon') = (0.229, 0.690)$ and
  $\alpha_Z = 0.685$ with $\chi^2_{min} = 1.44$ for $6-3=3$ \DOF.}
\label{eps-zenith}
\end{figure}

In Fig.~\ref{eps-zenith} we show the allowed regions in the
$(\varepsilon,\varepsilon')$ parameter space for neutrino scattering
off $d$- and $u$-quarks, respectively.  The contours in
Fig.~\ref{eps-zenith} correspond to the allowed regions at 90, 95 and
99\,\%~\CL.  The best fit (indicated by the open circle) is obtained
for $(\epsilon_\nu^d, {\epsilon'}_\nu^d) = (0.251, 0.62$) and
$\alpha_Z = 0.819$ with $\chi^2_{min} = 1.10$ for neutrino scattering
off $d$-quarks and at $(\epsilon_\nu^u, {\epsilon'}_\nu^u) = (0.229,
0.690$) and $\alpha_Z = 0.685$ with $\chi^2_{min} = 1.44$ for neutrino
scattering off $u$-quarks (having $6-3=3$ \DOF\ in both cases).

\begin{figure}[ht]
  \vglue -2.5cm \vglue 1.5cm \hglue -1.5cm
  \epsfig{file=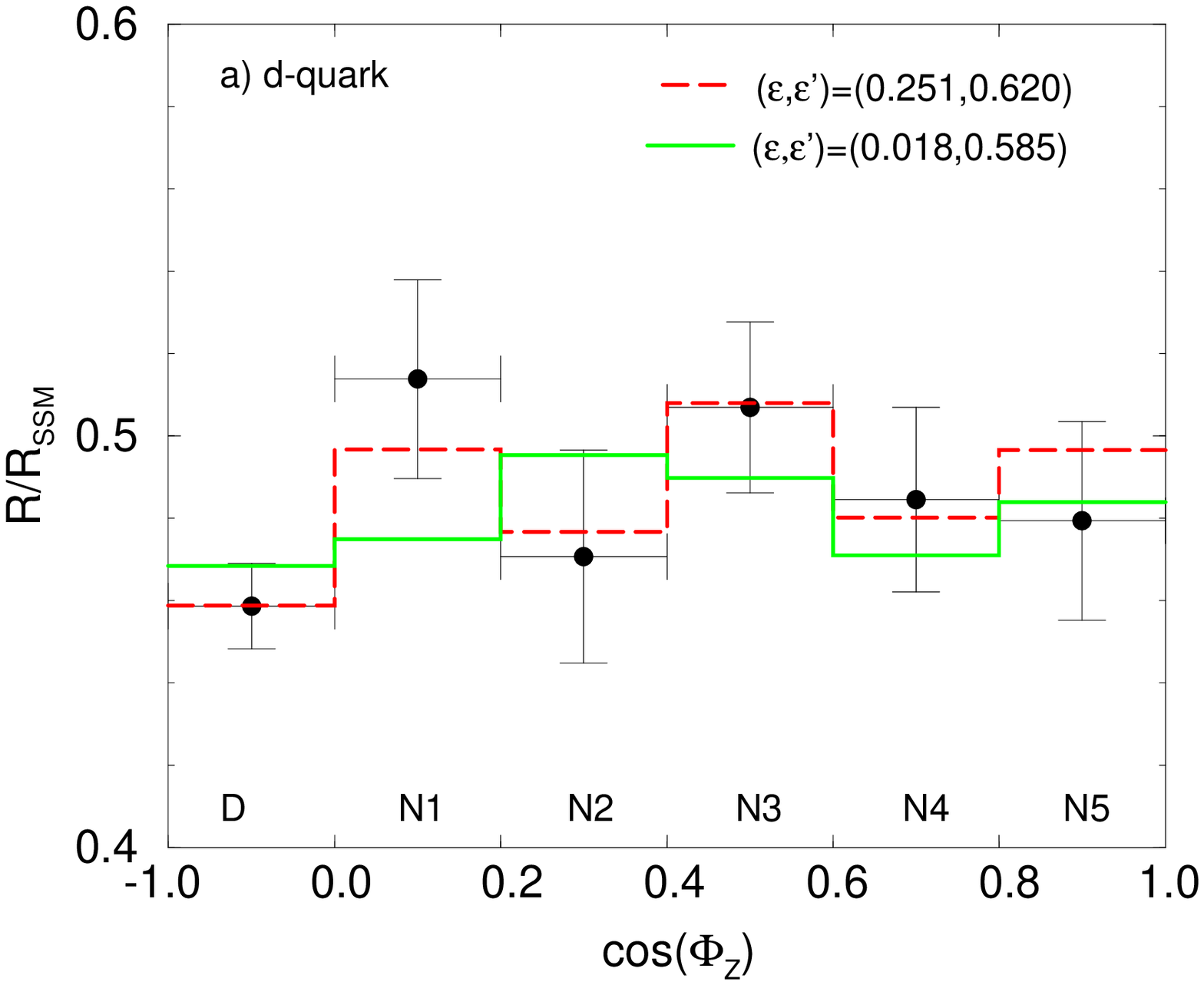,width=9.5cm} 
\vglue -7.8cm \hglue 7.0 cm
  \epsfig{file=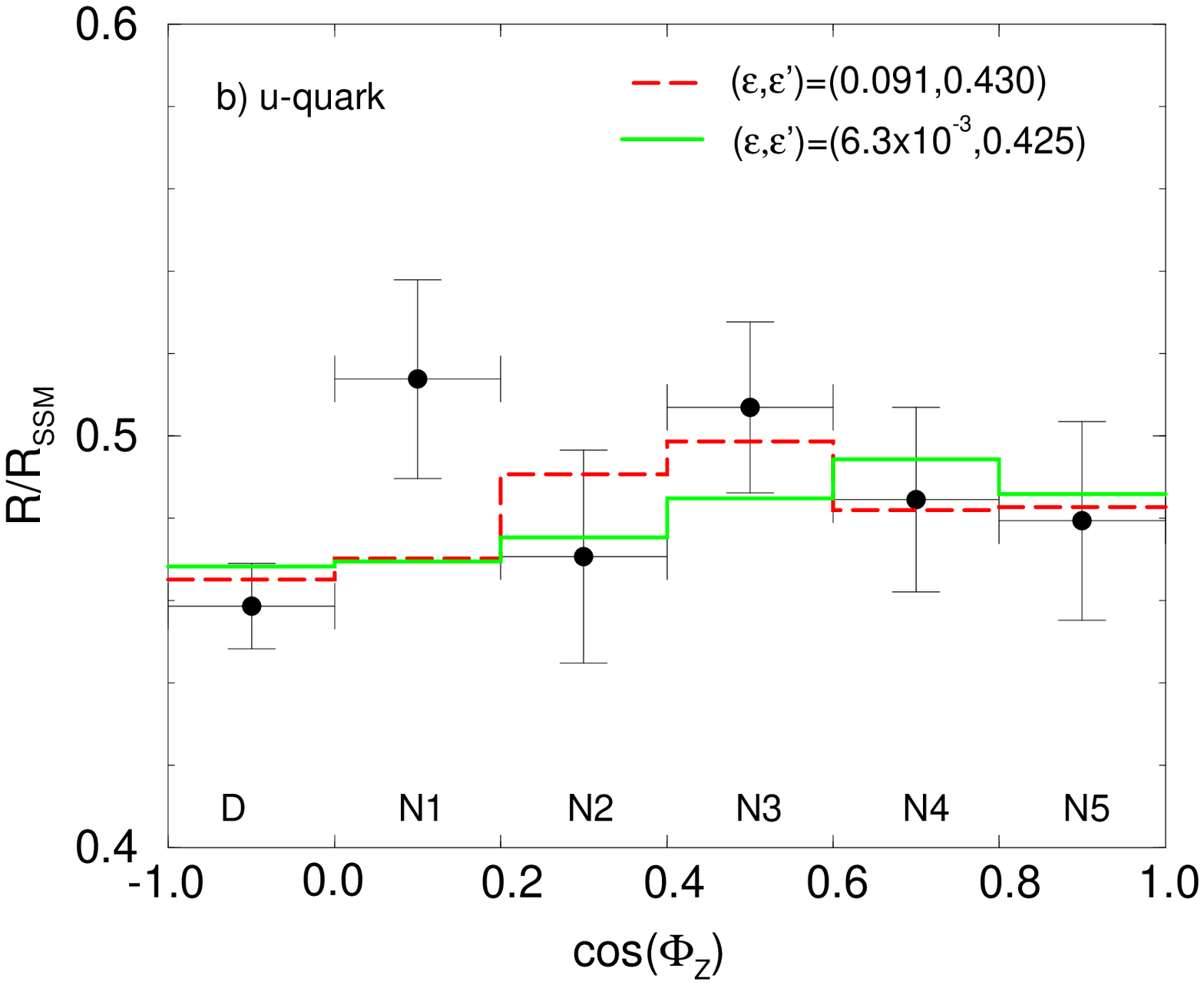,width=9.5cm}
\caption{\noindent
  Expected the zenith angle dependence with the our best fit values of
  $(\varepsilon, \varepsilon')$ determined by the SK Zenith angle only
  as well as the combined analysis for (a) $d$-quarks and (b)
  $u$-quarks.}  
\vglue -0.5cm
\label{zenith}
\end{figure}

Finally, in Fig.~\ref{zenith} we show the expected zenith angle
distributions for SuperKamiokande using the values of $(\epsilon,
\epsilon')$ determined by the best fit.  For comparison, we also
present in this figure the expected zenith angle distributions for the
best fit values of $(\epsilon, \epsilon')$ found in the combined
analysis (that will be discussed in section~\ref{combined}).

\subsection{Recoil Electron Spectrum}

We also consider the measurements of the recoil electron spectrum by
SuperKamiokande~\cite{Suzuki}. The available data, after 825 days of
operation, are divided into 18 bins. 17 of these bins have a width of
0.5 MeV and are grouped into two bins for a super low energy analysis
with energies between 5.5 MeV and 6.5 MeV and 15 bins with energies
ranging from 6.5 MeV (the low energy limit) to 14 MeV.  The last bin
includes all the events with energies larger than 14 MeV.

Since the electron neutrino survival probability does not depend on
the neutrino energy in the NSNI scenario, the spectral distortion of
the recoil electrons from $^8$B neutrino due to the presence of a
$\nu_{\mu,\tau}$ component in the neutrino flux is expected to be very
small~\cite{JNB2} and therefore, even a relatively small spectral
distortion (such as the one expected in small mixing angle MSW
solution) could rule out this solution.

The $\chi^2$-function that characterizes the deviations of the
measured ($S^{obs}_i$) from the predicted ($S^{th}_i$) values for the
electron recoil spectrum therefore provides an important test of the
NSNI solution. It is defined as:
\beq
\chi^2_S(\varepsilon,\varepsilon',\alpha_S) = \sum_{i,j=1,...,18} 
\left(\alpha_S S^{th}_i(\varepsilon,\varepsilon') - S^{obs}_i \right)
\left[\sigma^2_S \right]^{-1}_{ij}
\left(\alpha_S S^{th}_j(\varepsilon,\varepsilon') - S^{obs}_j \right) \,,
\eeq
where the error matrix (squared)
\beq 
\left[\sigma_S^2 \right]_{ij} =
 \delta_{ij} \left[\sigma^2_i(stat.) + \sigma^2_i(uncorr.)\right] + 
 \sigma_i(corr.) \, \sigma_j(corr.) + 
 \sigma_i(theor.) \, \sigma_j(theor.)
\eeq
includes statistical [$\sigma_i(stat.)$] and systematic experimental
errors (including both the uncorrelated [$\sigma_i(uncorr.)$] and the
correlated [$\sigma_i(corr.)$] contributions) as well as the
theoretical errors [$\sigma_i(theor.)$] (see Refs.~\cite{BKS,GHPV} for
more details.). Again, as in the analysis for the zenith angle
dependence, we introduce an overall normalization factor $\alpha_S$,
which is taken as a free parameter and determined from the fit, in
order to avoid over-counting the data on the total event rate.
Fitting the present data to our scenario we obtain $\chi^2_{min}=20.0$
for $18-1=17$~\DOF, which is still acceptable at the 27\,\%~\CL.

\subsection{Seasonal Variations}

The earth matter effects on neutrino flavor transitions induce a
seasonal variation of the data (beyond the expected variation of the
solar neutrino flux due to the eccentricity of the earth's orbit) due
to the variation of the day and night time during the year.  Since
these variations can be relevant to other neutrino oscillation
scenarios~\cite{seasonal}, a positive signal could help to distinguish
the various solutions and it is worthwhile to analyze the effects of
such a variation in the NSNI scenario.

The present SK solar neutrino data do not provide any conclusive
evidence in favor of such a variation, but indicate only that the
variation seems to be larger for recoil electron energies above 11.5
MeV.  In our scenario, however, we do not expect any correlation
between the seasonal variation and the recoil electron energies, since
the electron neutrino survival probability does not depend on the
neutrino energy.  Therefore any range of parameters that leads to a
considerable seasonal modulation for energies above 11.5 MeV is
disfavored by the data for lower energies.  However, for the range of
parameters ($\varepsilon, \varepsilon'$) that can solve the solar
neutrino problem, earth regeneration effects are never strong enough
to induce a significant seasonal variation.  Hence taking into account
the data on seasonal variations neither changes the shape of the
allowed region, nor the best fit points.

\subsection{Combined Analysis}
\label{combined}

\begin{figure}[ht]
\vglue -1.5cm
\vglue 1.5cm \hglue -1.5cm
\epsfig{file=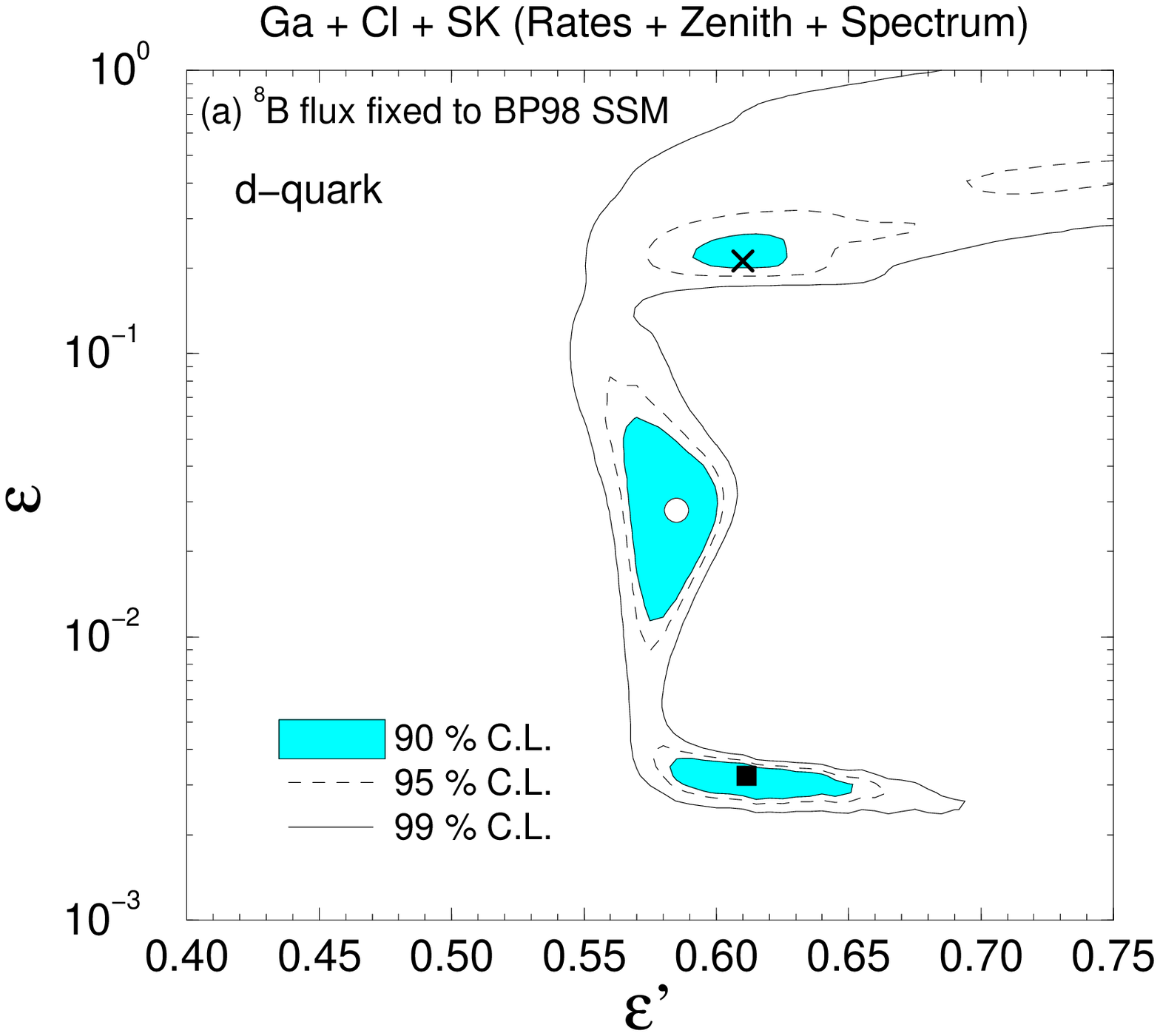,width=9.5cm}
\vglue -7.95cm \hglue 7.0 cm 
\epsfig{file=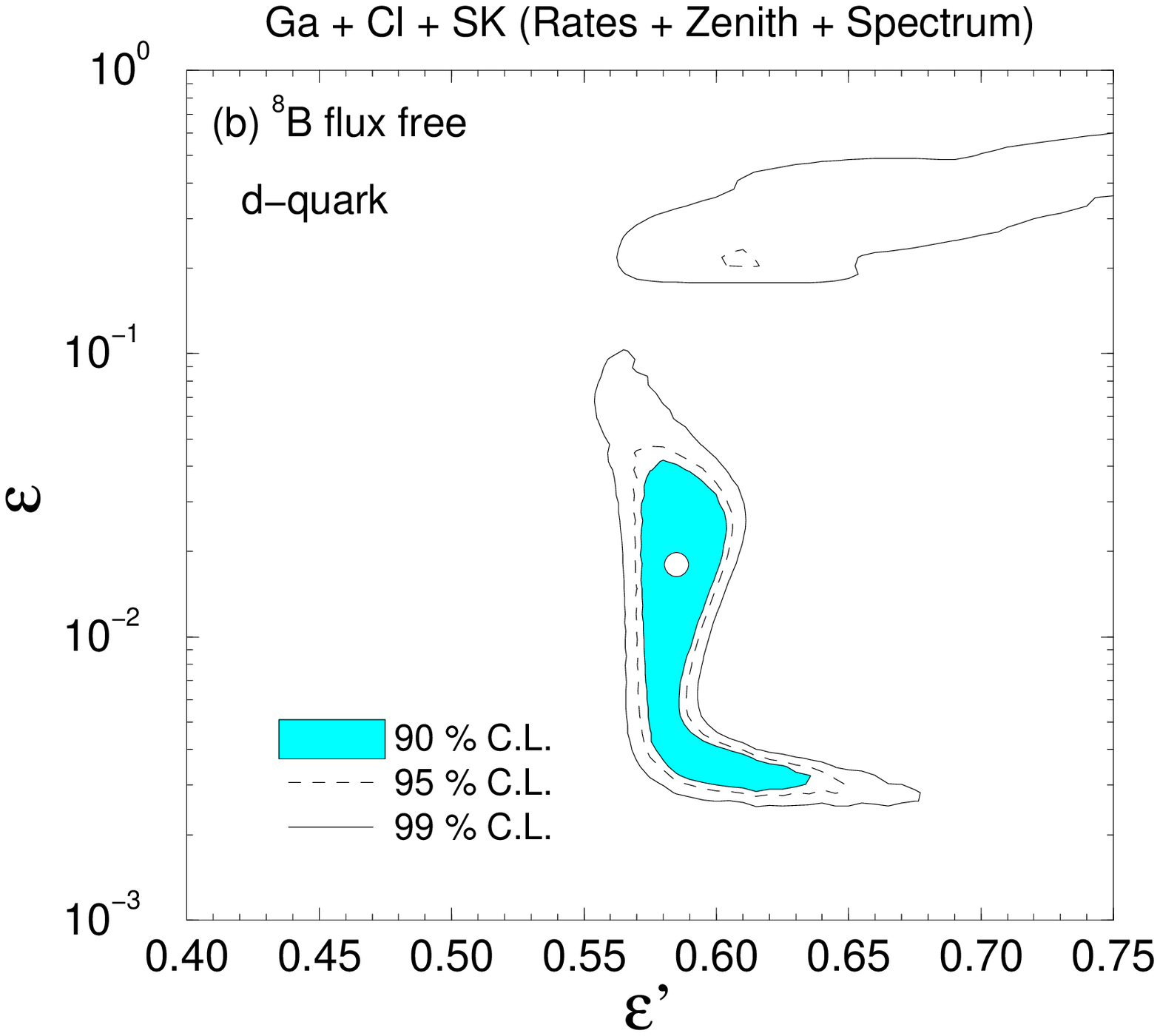,width=9.5cm}
\caption{\noindent 
  The allowed region for $\varepsilon=\epsilon_\nu^d$ and
  $\varepsilon' = {\epsilon'}_\nu^d$ obtained by the combined analysis
  using 4 rates + 6 zenith angle bins + 18 spectrum bins for
  non-standard neutrino interactions with $d$-quarks. (a) Fixing
  $f_B=1$ the best fit (indicated by the open circle) is obtained for
  $(\varepsilon, \varepsilon') = (0.028, 0.585)$ with $\chi^2_{min} =
  29.05$ for $28-4=24$ \DOF. There are two additional (local) $\chi^2$
  minima at $(\varepsilon, \varepsilon') = (0.0033, 0.610)$ with
  $\chi^2 = 29.40$ (indicated by the solid square) and $(\varepsilon,
  \varepsilon') = (0.21, 0.61)$ with $\chi^2 = 33.1$ (indicated by the
  cross). (b) Same as in (a) but allowing a free $f_B$. The best fit
  (indicated by the open circle) is obtained for $(\varepsilon,
  \varepsilon') = (0.018, 0.585)$ and $f_B = 1.38$ with $\chi^2_{min}
  = 26.62$ for $28-5=23$ \DOF.}  
\vglue 0.5cm
\label{eps-combined-d}
\end{figure}

\begin{figure}[ht]
\vspace{-2.3cm}
\vglue 1.5cm \hglue -1.5cm
\epsfig{file=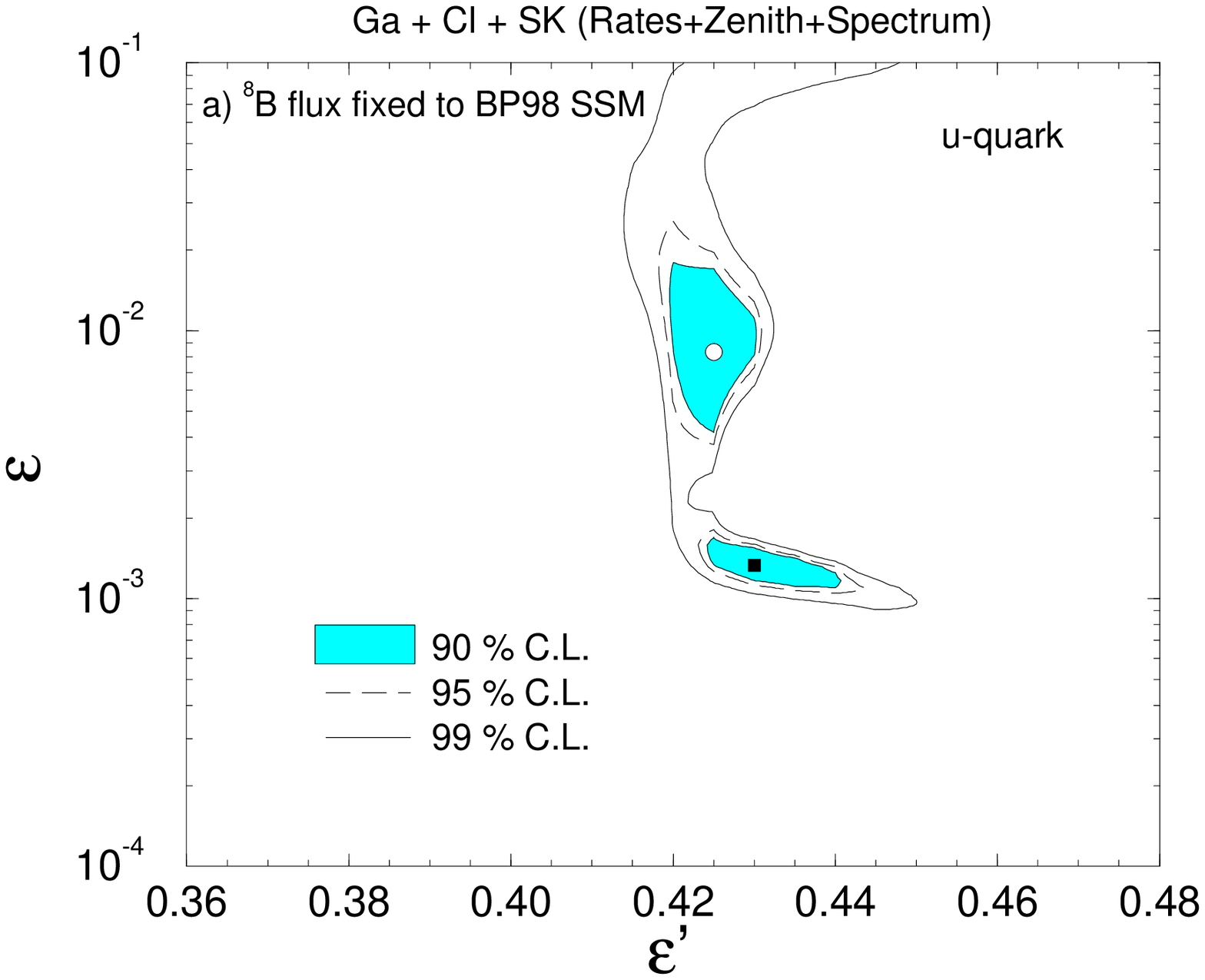,width=9.1cm}
\vglue -7.5cm \hglue 7.0 cm 
\epsfig{file=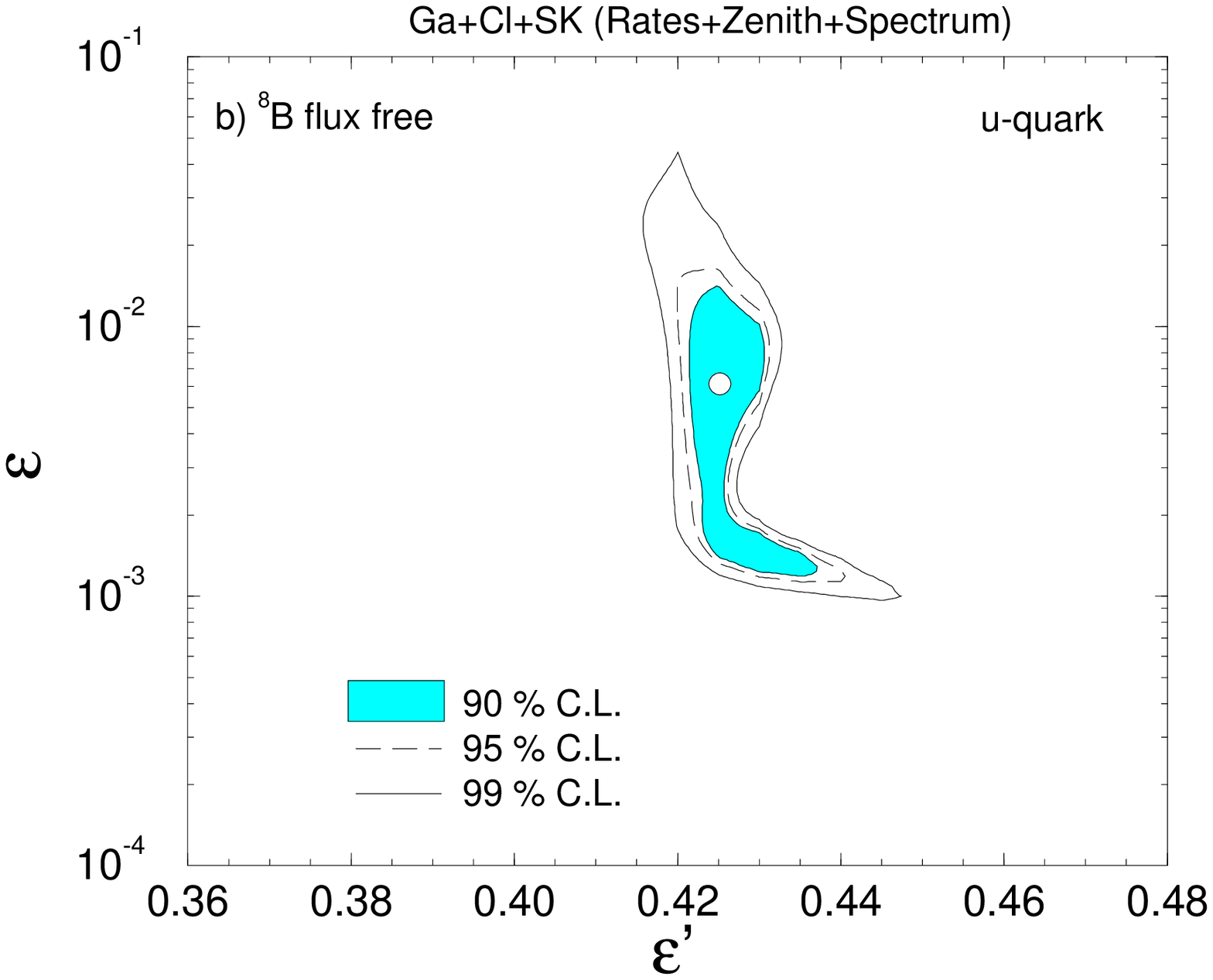,width=9.1cm}
\caption{\noindent
  Same as in Fig.~\ref{eps-combined-d} but for $u$-quarks.  (a) Fixing
  $f_B=1$ the best fit (indicated by the open circle) is obtained for
  $(\varepsilon, \varepsilon') = (0.0083, 0.425)$ with $\chi^2_{min} =
  28.45$ for $28-4=24$ \DOF. A second (local) $\chi^2$ minimum is
  found at $(\varepsilon, \varepsilon') = (0.0013, 0.430)$ with
  $\chi^2 = 30.27$ (indicated by the solid square) (b) Same as in (a)
  but allowing a free $f_B$. The best fit (indicated by the open
  circle) is obtained for $(\varepsilon, \varepsilon') = (0.0063,
  0.426)$ and $f_B = 1.34$ with $\chi^2_{min} = 26.59$ for $28-5=23$
  \DOF.}  
\vglue -0.5cm
\label{eps-combined-u}
\end{figure}

Our final result is the fit derived from the combined analysis of all
presently available solar neutrino data.  In Fig.~\ref{eps-combined-d}
and Fig.~\ref{eps-combined-u} we show the allowed regions for
$(\epsilon_\nu^d, {\epsilon'}_\nu^d)$ and $(\epsilon_\nu^u,
{\epsilon'}_\nu^u)$, respectively, using both the results from the
total rates from the Chlorine, GALLEX, SAGE and SuperKamiokande solar
neutrino experiments together with the 6 bins from the SuperKamiokande
zenith angle data discussed previously.  Although adding the spectral
information to our analysis does not change the shape of allowed
regions nor the best fit points, it is included in order to determine
the quality of the global fit.  However, we do not take into account
the seasonal variation in our combined $\chi^2$ analysis, since the
effect is negligible.

For neutrino scattering off $d$-quarks the best fit for the combined
data is obtained for
\beq
\epsilon_\nu^d    = 0.028 \andthis
{\epsilon'}_\nu^d = 0.585
\eeq
with $\chi^2_{min} = 29.05$ for $28-4=24$ \DOF, corresponding to a
solution at the 22\,\%~\CL\ (see Fig.~\ref{eps-combined-d}a).
Allowing $f_B \neq 1$, the best fit is found at $(\epsilon_\nu^d,
{\epsilon'}_\nu^d) = (0.018, 0.585)$ and $f_B = 1.38$ with
$\chi^2_{min} = 26.62$ for $28-5=23$ \DOF, corresponding to a solution
at the 27\,\%~\CL\ (see Fig.~\ref{eps-combined-d}b). For neutrino
scattering off $u$-quarks the best fit for the combined data is
obtained for
\beq
\epsilon_\nu^u    = 0.0083 \andthis
{\epsilon'}_\nu^u = 0.425
\eeq
with $\chi^2_{min} = 28.45$ for $28-4=24$ \DOF\ corresponding to a
solution at the 24\,\%~\CL\ (see Fig.~\ref{eps-combined-u}a).
Allowing $f_B\neq 1$, the best fit is obtained for $(\epsilon_\nu^u,
{\epsilon'}_\nu^u) = (0.0063, 0.426)$ and $f_B = 1.34$ with
$\chi^2_{min} = 26.59$ for $28-5=23$ \DOF, corresponding to a solution
at the 27\,\%~\CL\ (see Fig.~\ref{eps-combined-u}b).  These results
have to be compared with the fit for standard model neutrinos, that do
not oscillate (where the \CL\ is smaller than $10^{-7}$), as well as
to the standard solutions of the solar neutrino problem in terms of
usual neutrino oscillations (36\,\%~\CL)~\cite{BKS,GHPV}.

Finally, in Fig.~\ref{zenith} we show the expected zenith angle
distributions for SuperKamiokande using the best fitted values of
$(\epsilon, \epsilon')$ from the combined analysis.

\section{Phenomenological constraints on $\epsilon$ and $\epsilon'$}
\label{constraints}

In this section we investigate whether the allowed regions for the
parameters $\epsilon_{\nu_\ell}^f$ and ${\epsilon'}_{\nu_\ell}^f$ are
at all phenomenologically viable. The analysis of non-standard
neutrino interactions that could be relevant for the solar neutrino
problem is similar to the discussions in Refs.~\cite{BG,BGP}, where
the possibility that FCNI explain the LSND results~\cite{LSND,BG} or
the atmospheric neutrino anomaly~\cite{AN,BGP} was discussed.

Generically, extensions of the Standard Model include additional
fields that can induce new interactions: A heavy boson ${\cal B}$ that
couples weakly to some fermion bilinears $B_{ij}$ with the trilinear
couplings $\lambda_{ij}$, where $i, j = 1, 2, 3$ refer to fermion
generations, induces the four fermion operator $B_{ij}^\dagger B_{kl}$
at tree-level. The effective coupling is given by
\beq \label{GN} 
G^{B^\dagger B}_N = {{\lambda_{ij}^*} \lambda_{kl} \over 
                     {4 \sqrt2 M^2_{\cal B} } } \,, 
\eeq
for energies well below the boson mass $M_{\cal B}$.  Thus, in terms
of the trilinear coupling $\lambda_{\alpha f}$ that describes the
coupling of some heavy boson ${\cal B}$ to $\nu_\alpha$
($\alpha=e,\mu,\tau$) and a charged fermion $f=u,d,e$ the effective
parameters in~(\ref{defeps}) are given by
\beq \label{geneps}
\epsilon_{\nu_\ell}^f =    {\lambda_{\ell f}^* \lambda_{e f} 
                     \over  4 \sqrt2 M^2_{\cal B} \, G_F }
\andthis
{\epsilon'}_{\nu_\ell}^f = {|\lambda_{\ell f}|^2 - |\lambda_{e f}|^2 
                     \over  4 \sqrt2 M^2_{\cal B} \, G_F } \,.
\eeq

Since any viable extension of the Standard Model has to contain the SM
gauge symmetry, the effective theory approach presented in
Refs.~\cite{BG,BGP} is completely sufficient to describe any New
Physics effect for the energy scales typical to present neutrino
oscillation experiments. Even though the effective theory obviously
does not contain all the information inherent in the full high-energy
theory, the parameters of the effective theory are all of what is
accessible at low energies, when the ``heavy degrees of freedom'' are
integrated out.

The crucial point for our analysis is the following: Since the SM
neutrinos are components of $SU(2)_L$ doublets, the same trilinear
couplings $\lambda_{\alpha f}$ that give rise to non-zero
$\epsilon_{\nu_\ell}^f$ or ${\epsilon'}_{\nu_\ell}^f$ also induce
other four-fermion operators.  These operators involve the $SU(2)_L$
partners of the neutrinos, i.e.  the charged leptons, and can be used
to constrain the relevant couplings.

Noting that Lorentz invariance implies that any fermionic bilinear
$B_{ij}$ can couple to either a scalar~($\calS$) or a vector~($\calV$)
boson it is straightforward to write down all gauge invariant
trilinear couplings between the bilinears (that contain SM fermions)
and arbitrary bosons $\calS$ and $\calV$ that might appear in a
generic extension of the Standard Model (see Tabs.~1--3 of
Ref.~\cite{BGP}). From these couplings one then obtains all the
effective four fermion operators relevant to the solution to the solar
neutrino problem in terms of NSNI as well as the $SU(2)_L$-related
operators that are used to constrain their effective couplings.  (We
do not consider here operators that violate total lepton number which
can be induced if there is mixing between the intermediate
bosons~\cite{LNV}.)

While we refer the reader to Refs.~\cite{BG,BGP} for the details of
this model-independent approach, we present here two explicit examples
relevant to solar neutrinos to demonstrate how $SU(2)_L$ related
processes can be used to constrain the parameters
$\epsilon_{\nu_\ell}^f$ or ${\epsilon'}_{\nu_\ell}^f$. First, consider
the bilinear $\bar L f_R$ (where $L$ denotes the lepton doublet and
$f=e, u, d$) that couples via a scalar doublet to its hermitian
conjugate $\lbar{f_R} L$. In terms of the component fields the
effective interaction is
\beqs
&~&
{\lambda^*_{\alpha f} \lambda_{\beta f} \over M_1^2}
 (\lbar{\nu_\alpha} f_R)\,(\lbar{f_R} \nu_\beta) + 
{\lambda^*_{\alpha f} \lambda_{\beta f} \over M_2^2}
 (\lbar{l_\alpha} f_R)\,(\lbar{f_R} l_\beta) ~ = \nonumber \\
-&~&
{\lambda^*_{\alpha f} \lambda_{\beta f} \over 2 M_1^2}
 (\lbar{\nu_\alpha} \gamma^\mu \nu_\beta)\,(\lbar{f_R} \gamma_\mu f_R)  
-{\lambda^*_{\alpha f} \lambda_{\beta f} \over 2 M_2^2}
  (\lbar{l_\alpha} \gamma^\mu l_\beta)\,(\lbar{f_R} \gamma_\mu f_R) \,, 
\label{fierzed}
\eeqs
where $l_\alpha = e_L, \mu_L, \tau_L$ for $\alpha = e, \mu, \tau$.
$\lambda_{\alpha f}$ is the trilinear coupling of $\lbar{L_\alpha}
f_R$ to the scalar doublet and $M_{1,2}$ denote the masses of its
$SU(2)_L$ components.  The important point is that the scalar doublet
exchange not only gives rise to the four-Fermi operator ${\cal
  O}^f_\nu$ in~(\ref{Onu}) (with $(V-A) \otimes (V+A)$ structure), but
also produces the $SU(2)_L$ related operator
\beq \label{Oell}
{\cal O}_l^f \equiv 
 (\lbar{l_\alpha} \, l_\beta) \, (\bar f \, f) \,,
\eeq
which has the same Lorentz structure as ${\cal O}_\nu^f$, {\sl with 
the neutrinos replaced} by their charged lepton partners. Moreover,
the effective coupling of ${\cal O}^f_l$, that we denote by
$G_{\alpha \beta}^f$, is related to $G_{\nu_\alpha \nu_\beta}^f$ by
\beq \label{Gratio}
G_{\nu_\alpha \nu_\beta}^f = G_{\alpha \beta}^f \, {M_1^2 \over M_2^2}\,.
\eeq

Constructing all the relevant four fermion operators that are induced
by the couplings between the bilinears listed in Tabs.~1--3 of
Ref.~\cite{BGP}, one finds that in general ${\cal O}^f_l$ is generated
together with ${\cal O}^{f'}_\nu$. Here $f'$ can be different from $f$
only for interactions with quarks, that is in some cases ${\cal
O}^u_l$ (${\cal O}^d_l$) is generated together with ${\cal O}^d_\nu$
(${\cal O}^u_\nu$). The leptonic operator ${\cal O}^e_l$ is always
generated together with ${\cal O}^e_\nu$ unless the interaction is
mediated by an intermediate scalar $SU(2)_L$ singlet that couples to
\beq
(L_\ell L_e)_s = 
{1 \over \sqrt2} (\lbar{\nu_\ell^c} e_L - \lbar{\ell_L^c} \nu_e) \,.
\eeq
Note that the singlet only couples between two different flavors,
since the coupling has to be antisymmetric in flavor space.
Consequently a singlet that couples to the bilinear $(L_\ell L_e)_s$
cannot induce a non-zero $\epsilon_{\nu_\ell}^e$. The fact that the
resulting four fermion operators only mediate FDNI is true because for
the solar neutrinos we only care about $\nu_e \to \nu_\ell$
transitions. (For atmospheric neutrinos also $\nu_\mu \to \nu_\tau$
transitions induced by \NSNI\ with the electrons are of interest. In
this case the coupling of $(L_\mu L_e)_s$ to $(L_\tau L_e)_s^\dagger$
via singlet exchange inducing FCNI is possible~\cite{BGP}.)

The effective interactions that are mediated by a scalar singlet of
mass $M$ that couples to $(L_\ell L_e)_s$ with the elementary coupling
$\lambda_{\ell e}$ are given by
\beqs
&~&
{|\lambda_{\ell e}|^2 \over M^2} 
\left[
 (\lbar{e_L} \nu_\ell^c)\,(\lbar{\nu_\ell^c} e_L) 
-(\lbar{e_L} \nu_\ell^c)\,(\lbar{\ell_L^c} \nu_e) 
+(\lbar{\nu_e} \ell_L^c)\,(\lbar{\ell_L^c} \nu_e) 
-(\lbar{\nu_e} \ell_L^c)\,(\lbar{\nu_\ell^c} e_L) 
\right] = \nonumber \\
&~& 
{|\lambda_{\ell e}|^2 \over M^2}
\left[
 (\lbar{e_L} \gamma^\mu e_L)\,(\lbar{\nu_\ell} \gamma_\mu \nu_\ell)
-(\lbar{e_L} \gamma^\mu \nu_e)\,(\lbar{\nu_\ell} \gamma_\mu \ell_L) 
+(\lbar{\nu_e} \gamma^\mu \nu_e)\,(\lbar{\ell_L} \gamma_\mu \ell_L)
-(\lbar{\nu_e} \gamma^\mu e_L)\,(\lbar{\ell_L} \gamma_\mu \nu_\ell)
\right] \,, 
\label{fierzed2}
\eeqs
where we used a Fierz transformation and the identity $\lbar{A^c}
\gamma^\mu B^c = -\lbar{B} \gamma^\mu A$ to obtain~(\ref{fierzed2}).
One can see that in this case ${\cal O}^e_\nu$ is generated together
with three more operators that have the same effective coupling (up to
a sign).  However, unlike for the case of intermediate doublets (or
triplets), all these operators involve two charged leptons and two
neutrinos.
 
\subsection{Experimental constraints}

\subsubsection{Flavor changing neutrino interactions}

There is no experimental evidence for any non-vanishing $G_{e
  \ell}^f$\,. Therefore, whenever ${\cal O}^f_l$ is generated together
with ${\cal O}^f_\nu$, one can use the upper bounds on $G_{e \ell}^f$
to derive constraints on $G_{\nu_e \nu_\ell}^f$.  The most stringent
constraints on $G_{e \ell}^e$ are due to the upper bounds on
$\MuToEee$ and $\TauToEee$~\cite{Bliss,PDG}:
\beqs
\BR(\MuToEee)  &<& 1.0 \times 10^{-12} \,, \label{MuEbound} \\
\BR(\TauToEee) &<& 2.9 \times 10^{-6}  \,. \label{TauEbound}
\eeqs
Normalizing the above bounds to the measured rates of the related lepton
flavor conserving decays, $\BR(\mu^- \to e^- \, \bar\nu_e \,
\nu_\mu) \approx 100\,\%$ and $\BR(\tau^- \to e^- \, \bar\nu_e \,
\nu_\tau) = 0.18$~\cite{PDG}, we obtain
\beqs
\epsilon_{\mu}^e \equiv G_{e \mu}^e / G_F  &<& 
 1.0 \times 10^{-6} \,, \label{GeBoundMu} \\
\epsilon_{\tau}^e \equiv G_{e \tau}^e / G_F &<& 
 4.2 \times 10^{-3} \,. \label{GeBoundTau}
\eeqs
Note that the bounds on $\epsilon_\ell^f$ do only coincide with those
for $\epsilon_{\nu_\ell}^f$ in the $SU(2)$ symmetric limit. We will
comment on possible relaxations due to $SU(2)_L$ breaking effects
later in section~\ref{breaking}.

To constrain $G_{e \mu}^q$ we use the upper bounds on $\mu \to e$
conversion from muon scattering off nuclei~\cite{PDG},
\beqs
\MuToEonTi ~&<&~ 4.3 \times 10^{-12} \,, \label{TiBound} \\
\MuToEonPb ~&<&~ 4.6 \times 10^{-11} \,, \label{PbBound} \\
\MuToEonS  ~&<&~ 7   \times 10^{-11} \,, \label{SBound}  
\eeqs
concluding that
\beq \label{GqBoundMu}
\epsilon_\mu^q \equiv G_{e \mu}^q / G_F \lsim 10^{-5} \,.
\eeq
is a conservative upper bound irrespective of the inherent hadronic
uncertainties for such an estimate.

To constrain $G_{e \tau}^q$ we may use the upper bounds on various
semi-hadronic tau decays that violate lepton flavor~\cite{Bliss,PDG}:
\beqs
\BR(\TauToEPi)   ~&<&~ 3.7 \times 10^{-6} \,, \label{PiBound}  \\
\BR(\TauToERho)  ~&<&~ 2.0 \times 10^{-6} \,, \label{RhoBound} \\
\BR(\TauToEEta)  ~&<&~ 8.2 \times 10^{-6} \,, \label{EtaBound} \\
\BR(\TauToEPiPi) ~&<&~ 1.9 \times 10^{-6} \,. \label{PiPiBound} 
\eeqs
Let us first consider the tau decays into $\pi^0$ and $\rho^0$.  Since
these mesons belong to an isospin triplet we can use the isospin
symmetry to normalize the above bounds~(\ref{PiBound})
and~(\ref{RhoBound}) by the measured rates of related lepton flavor
conserving decays.  Using $\BR(\tau^- \to \nu_\tau \pi^-) =
0.11$~\cite{PDG} and $\BR(\tau^- \to \nu_\tau \rho^-) =
0.22$~\cite{Don,PDG} we obtain
\beq \label{GqBound}
G_{e \tau}^q(\pi)  < 8.2 \times 10^{-3} \, G_F \,, \andthis
G_{e \tau}^q(\rho) < 4.2 \times 10^{-3} \, G_F \,. 
\eeq
Since the $\pi$ ($\rho$) is a pseudoscalar (vector) meson
its decay probes the axial-vector (vector) part of the quark current.

In general, any semi-hadronic operator ${\cal O}^q_l$ can be
decomposed into an $I=0$ and an $I=1$ isospin component.  Only the
effective coupling of the latter can be constrained by the upper
bounds on the decays into final states with isovector mesons, like the
$\pi$ and the $\rho$.  If the resulting operator is dominated by the
$I=0$ component, the bounds in~(\ref{GqBound}) do not hold. But in
this case we can use the upper bound on $\BR(\TauToEEta)$
in~(\ref{EtaBound}).  Since the $\eta$ is an isosinglet, isospin
symmetry is of no use for the normalization.  However, we can estimate
the proper normalization using the relation between the $\eta$ and
$\pi$ hadronic matrix elements, which is just the ratio of the
respective decay constants, $f_\eta/f_\pi \simeq 1.3$~\cite{Don,PDG}.
Taking into account the phase space effects, we obtain
from~(\ref{EtaBound}) that
\beq \label{GqBoundEta}
G_{e \tau}^q(\eta) < 1.1 \times 10^{-2} \, G_F \,.
\eeq
Since the $\eta$ is a pseudoscalar meson its decay probes the
axial-vector part of the $I=0$ component of the quark current, while
the neutrino propagation is only affected by the vector part.  As we
have already mentioned, for any single chiral New Physics contribution
the vector and axial-vector parts have the same magnitude and we can
use~(\ref{GqBoundEta}) to constrain the isosinglet component of ${\cal
O}_l^q$. In case there are several contributions, whose axial-vector
parts cancel each other~\cite{BGP}, the $I=0$ component could still be
constrained by the upper bound on $\BR(\TauToEPiPi)$
in~(\ref{PiPiBound}). While the calculation of the rate is uncertain
due to our ignorance of the spectra and the decay constants of the
isosinglet scalar resonances, we expect that the normalization will be
similar to that of the $\pi$, $\rho$ and $\eta$ discussed
before. Finally we note that the decay $\tau^- \to e^- \, \omega$
would be ideal to constrain the $I=0$ vector part, but at present no
upper bound on its rate is available.

While one can always fine-tune some parameters in order to avoid our
bounds, our basic assumption is that this is not the case.  Thus
from~(\ref{GqBound}) and~(\ref{GqBoundEta}) we conclude that
\beq \label{eps_tau}
\epsilon_\tau^q \equiv G_{e \tau}^q / G_F \lsim 10^{-2} \,.
\eeq

\subsubsection{Flavor diagonal neutrino interactions}

So far we have only discussed the upper bounds on FCNI.  However, if
the neutrinos are massless then in addition to the FCNI that induce an
off-diagonal term in the effective neutrino mass matrix, also
non-universal flavor diagonal interactions are needed to generate the
required splitting between the diagonal terms.

In general any operator that induces such FDNI is related to other
lepton flavor conserving operators, that give additional contributions
to SM allowed processes, and therefore violate the lepton universality
of the SM. Then the upper bounds on lepton universality violation can
be used to constrain these operators.  Using the relation to the
operators that induce the FDNI one may also constrain the latter.

As we mentioned already for massless neutrinos only a non-zero
${\epsilon'}_{\nu_\ell}^q$ ($q=u,d$) can lead to a resonance effect,
while FDNI that allow for scattering off electrons alone are
insufficient to solve the solar neutrino problem. Therefore we only
need to discuss the effective flavor diagonal operator
\beq \label{Onuq}
{\cal O}_\nu^q \equiv (\lbar{\nu_\alpha} \gamma_\mu \nu_\alpha) \, 
                      (\lbar q \gamma^\mu q) \,,
\eeq
where $\alpha=e, \mu, \tau$ and $q=u_{L,R}, d_{L,R}$.

It is easy to check that for FDNI induced by heavy boson exchange
${\cal O}_\nu^q$ is always induced together with
\beq \label{Oellq}
{\cal O}_l^q \equiv (\lbar{l_\alpha} \gamma_\mu l_\alpha) \, 
                       (\lbar{q'} \gamma^\mu q') \,,
\eeq
where $q'=u_{L,R}, d_{L,R}$ can be different from $q$.  Moreover, 
intermediate scalar singlets and triplets (that couple to $Q L$) as
well as charged vector singlets and triplets (that couple to $\bar
Q L$) also give rise to
\beq \label{Oellnuq}
{\cal O}_{l \nu}^q \equiv 
   (\lbar{l_\alpha} \gamma_\mu \nu_\alpha) \, 
   (\lbar{q_L} \gamma^\mu q'_L) \,,
\eeq
where $q'=u, d$ for $q=d, u$. (See~(\ref{fierzed2}) as an example for
FDNI mediated by an intermediate scalar singlet.) Since ${\cal O}_{l
  \nu}^q$ induces an additional contribution to the SM weak decay
$\tau_L \to \pi \, \nu_\tau$ for $\alpha=\tau$ and to $\pi \to
l_\alpha \, \bar\nu_\alpha$ for $\alpha=e, \mu$, the relevant
effective coupling $G_{l \nu}^q$ can be constrained by the upper
bounds on lepton universality violation in semi-hadronic decays.  The
latter leads to a deviation of the parameters
\beqs \label{Rmue}
R_{e/\mu}^\pi &\equiv& \sqrt{{1 \over N}
{\Gamma(\pi^- \to e^- \, \bar\nu_e) \over 
 \Gamma(\pi^- \to \mu^- \, \bar\nu_\mu)}}  
\approx 1 + {G_{e \nu_e}^q - G_{\mu \nu_\mu}^q \over G_F} \\
\label{Rtaumu}
R_{\tau/\mu}^\pi &\equiv& \sqrt{{1 \over N}
{\Gamma(\tau^- \to \nu_\tau \, \pi^-) \over
 \Gamma(\pi^- \to \mu^- \, \bar\nu_\mu)}}  
\approx 1 + {G_{\tau \nu_\tau}^q - G_{\mu \nu_\mu}^q \over G_F} \\
R_{\tau/e}^W &\equiv& \sqrt{{1 \over N}
{\Gamma(W^- \to \tau \, \bar\nu_\tau) \over
 \Gamma(W^- \to e \, \bar\nu_e)}}  
\approx 1 + {G_{\tau \nu_\tau}^q - G_{e \nu_e}^q \over G_F} 
\eeqs
from unity. Here $N$ denotes a normalization factor, which is just the
ratio of the above two rates in the SM such that $R_{\alpha/\beta} =
1$ if $G_{l_\alpha \nu_\alpha}^q=G_{l_\beta \nu_\beta}^q$. In the
approximation we assume that $G_{l \nu}^q \ll G_F$.  From
the most recent experimental data~\cite{Pich,PDG} it follows that
\beqs 
R_{e/\mu}^\pi    &=& 1.0017 \pm 0.0015 \,, \\
R_{\tau/\mu}^\pi &=& 1.005 \pm 0.005 \,, \\
R_{\tau/e}^W     &=& 0.987 \pm 0.023 \,, 
\eeqs 
implying that
\beq \label{Gellnubound} 
{\epsilon'}_\ell^q \equiv 
 {G_{\ell \nu_\ell}^f - G_{e \nu_e}^q \over G_F}
\lsim 10^{-2} 
\eeq
is a conservative upper bound.  If ${\cal O}_{l \nu}^q$ is induced
together with ${\cal O}_\nu^q$, then in the $SU(2)_L$ symmetric limit
${\epsilon'}_\ell^q = {\epsilon'}_{\nu}^q$, but a modest relaxation
due to $SU(2)_L$ breaking effects is possible (see
section~\ref{breaking}).

It is essential to realize that not all New Physics operators that
induce the FDNI relevant to solar neutrinos are related to ${\cal
O}_{l \nu}^q$. For an intermediate $SU(2)_L$ scalar doublet (see
eq.~(\ref{fierzed}) for $f=q$) or a vector doublet (that couples to
$\lbar{q^c} L$) or a neutral vector singlet, only ${\cal O}_l^q$ is
induced together with ${\cal O}_\nu^q$.  In this case one may only use
the upper bound on $G_{l l}^q$ that are due to the constraints on
compositeness. The present data from $p \, \bar p \to e^+ e^-, \mu^+
\mu^- + X$~\cite{Abe,PDG} imply an upper limit on the scale of
compositeness $\Lambda(q q l_\alpha l_\alpha) \gsim 1.6$\,TeV, which
translates into
\beq \label{GqBound2}
G_{l_\alpha l_\alpha}^q \lsim 10^{-1} G_F
\eeq
as a conservative estimate for $\alpha=e, \mu$. (One-loop
contributions to the $Z$ width due to ${\cal O}_l^q$ lead to a similar
constraint on $G_{ee}^q$~\cite{GGN}.) However, no upper bound on
$\Lambda(q q \tau \tau)$ is available.

For a neutral vector singlet $G_{l_\alpha l_\alpha}^q = G_{\nu_\alpha
\nu_\alpha}^q$ and from~(\ref{defeps}) and~(\ref{GqBound2}) it follows
that
\beq \label{epsprimebound}
{\epsilon'}_{\nu_\mu}^q \lsim 10^{-1} \,
\eeq
while there is no model-independent bound on
${\epsilon'}_{\nu_\tau}^q$. For intermediate $SU(2)_L$ doublets the
bound in~(\ref{epsprimebound}) could be relaxed somewhat, since the
effective couplings of the relevant operators may differ due to
$SU(2)_L$ breaking effects, which we discuss next.

\subsection{Constraining $SU(2)_L$ breaking effects}
\label{breaking}

The excellent agreement between the SM predictions and the electroweak
precision data implies that $SU(2)_L$ breaking effects cannot be
large.  To show that the upper bounds on $G_{l l}^f$ (or $G_{l
\nu}^f$) translate into similar bounds for $G_{\nu \nu}^f$ if their
related operators stem from the same $SU(2)_L$ invariant coupling, we
recall from eq.~(\ref{Gratio}) that in general the ratio of the
couplings, $G_{\nu_\alpha \nu_\beta}^f / G_{\alpha \beta}^f$ (or
$G_{\nu_\alpha \nu_\beta}^f / G_{\alpha \nu_\beta}^f$), is given by
ratio $M_1^2 / M_2^2$.  Here $M_1$ and $M_2$ are the masses of the
particles belonging to the $SU(2)_L$ multiplet that mediate the
processes described by $G_{\alpha \beta}^f$ ($G_{\alpha \nu_\beta}^f$)
and $G_{\nu_\alpha \nu_\beta}^f$, respectively.  If $M_1 \ne M_2$ this
multiplet will contribute to the oblique parameters~\cite{obl} $S, U$
and, most importantly, $T$\,.  A fit to the most recent precision data
performed in Ref.~\cite{BGP} determined the maximally allowed ratio
$(M_1/M_2)^2_{\rm max}$ to be at most 6.8 (at 90\,\%~\CL) for
intermediate scalars. (Vector bosons in general are expected to have
even stronger bounds for the mass ratio). Consequently the upper
limits on the effective couplings $G_{\nu \nu}$ agree with those we
derived for the corresponding $G_{l l}^f$ (or $G_{l \nu}^f$) within an
order of magnitude even for maximal $SU(2)_L$ breaking. Thus, barring
fine-tuned cancellations. 
\beq
\epsilon^f_{\nu_\ell}    < 6.8 \, \epsilon^f_\ell \andthis
{\epsilon'}^f_{\nu_\ell} < 6.8 \, {\epsilon'}^f_\ell \,,
\eeq
at 90\,\%~\CL.

\section{Implications for Future Experiments}
\label{futureexperiments}

In this section we discuss how to test the solution to the solar
neutrino problem based on \NSNI\, in future neutrino experiments.

Let us consider first the possibility of obtaining stronger
constraints on New Physics from future laboratory experiments.  Our
phenomenological analysis shows that FCNI could only be large enough
to provide $\nu_e \to \nu_\tau$ transitions while,
model-independently, $\nu_e \to \nu_\mu$ transitions are irrelevant
for solar neutrinos. Even for $\nu_e \to \nu_\tau$ transitions the
required effective coupling has to be close to its current upper
bound, which we derived from limits on anomalous tau decays. Therefore
the solution to the solar neutrino problem studied in this paper could
be tested by the upcoming $B$-factories that are expected to improve
the present experimental bounds on several rare $\tau$ decays.  For
example, assuming an integrated luminosity of 30~fb$^{-1}$
(corresponding to $3 \times 10^7 \tau$~pairs) for the
BaBar~\cite{BaBar} experiment, the upper limits on the branching
ratios in~(\ref{PiBound})--(\ref{PiPiBound}) could be reduced by one
order of magnitude. This would decrease the bound on $\epsilon_\tau^q$
in~(\ref{eps_tau}) to a value close to the smallest possible best fit
values for~$\varepsilon$ (c.f.  Fig.~\ref{eps-combined-d} and
Fig.~\ref{eps-combined-u}) ruling out a large region of the parameter
space and making the NSNI solution increasingly fine-tuned.

\begin{figure}[ht]
\vglue 1.0cm \hglue -1.0cm
\epsfig{file=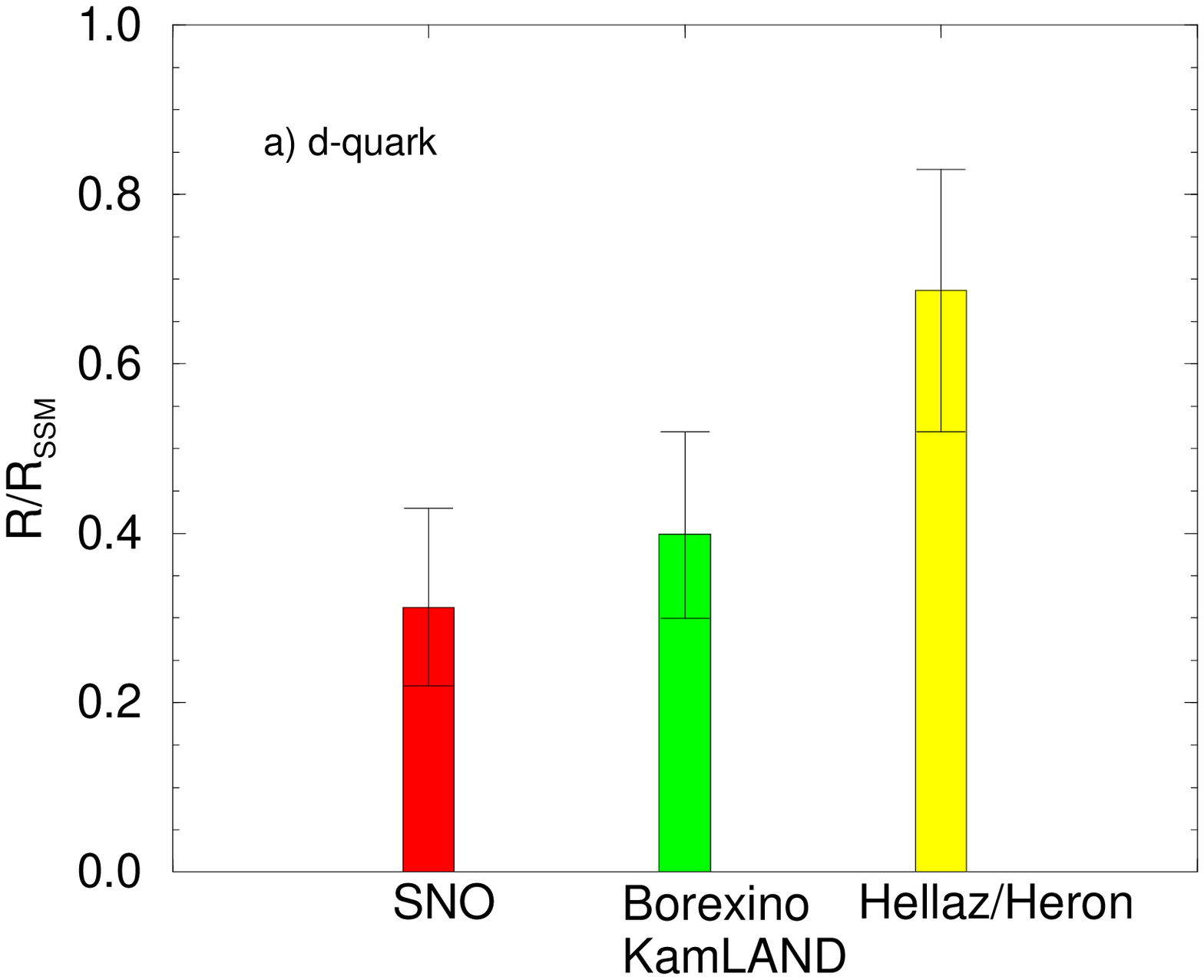,width=9.0cm}
\vglue -7.4cm \hglue 7.0 cm 
\epsfig{file=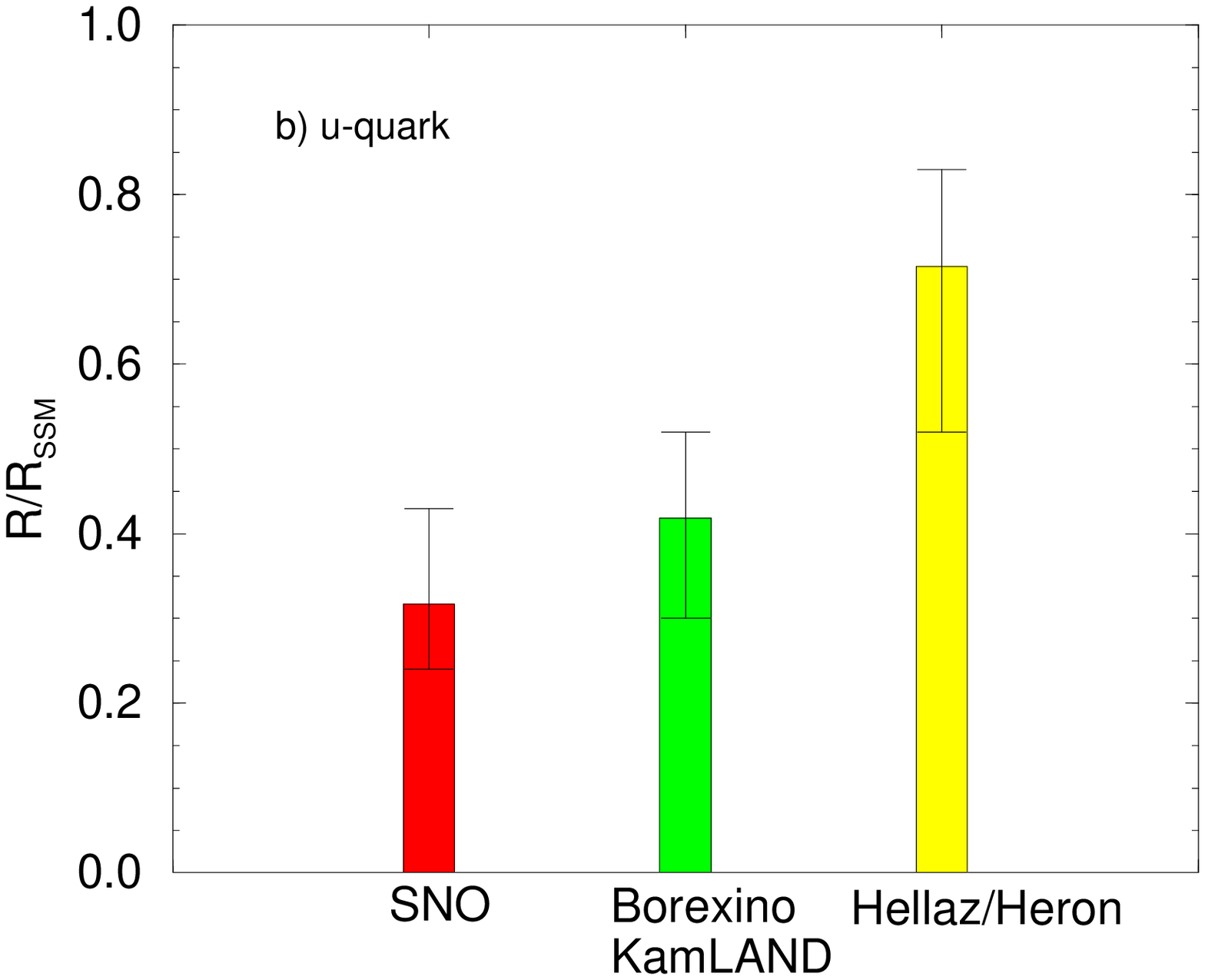,width=9.0cm}
\caption{\noindent
  The predicted ranges of the ratios of event rates for SNO, BOREXINO
  and HELLAZ/HERON to the corresponding event rates predicted from the
  SSM assuming that the neutrino conversion induced by FCNI as well as
  FDNI is the solution to the solar neutrino problem.  Histgrams
  indicate the predictions with the best fittes parameters whereas the
  error bars indicates the range determined by varying parameters
  within 95 \% C.L. regions for $(\varepsilon, \varepsilon')$ in
  Figs.~\ref{eps-combined-d}b and ~\ref{eps-combined-u}b.}  
\vglue0.5cm
\label{future}
\end{figure}

Next we consider the implications for future solar neutrino
experiments~\cite{SNO,BOREXINO,KamLAND,HELLAZ,HERON}. In Tab.~1 we
present the expected ranges for the event rates (normalized to the SSM
expectation in the absence of neutrino flavor transitions) of those
experiments, if the solar neutrino problem is explained by NSNI. In
Fig.~\ref{future} the predicted rates are presented graphically. The
ranges correspond to the 95\,\%~\CL\ regions for $(\varepsilon,
\varepsilon')$ in Figs.~\ref{eps-combined-d}b and
~\ref{eps-combined-u}b. As before we use the BP98 SSM predictions for
the initial neutrino fluxes and the survival probability in
eq.~(\ref{day}) to compute the expected rates for each of the five
detectors.  Specifically, there are three types of detectors: (a) The
Sudbury Neutrino Observatory (SNO)~\cite{SNO}, which is measuring the
$^8$B neutrino charged current (CC) rate, (b) the
BOREXINO~\cite{BOREXINO} and KamLAND~\cite{KamLAND} experiments that
are designed to observe the $^7$Be neutrino signal and (c) the
HELLAZ~\cite{HELLAZ} and HERON~\cite{HERON} experiments dedicated to a
precise measurement of the low-energy $pp$ neutrino flux.

\vspace{0.5cm}
\begin{center} 
Tab.~1: Future solar neutrino experiments and their rates predicted by
the NSNI solution \\
\vspace{0.5cm}
\begin{tabular}{| c | c | c | c |} 
\hline  
Experiment & Start of operation & Main neutrino source & 
Rate predicted by NSNI \cr
\hline
\hline
SNO        & 1999               & $^8$B    & $0.22-0.43$ \cr
\hline
BOREXINO   & 2001               & $^7$Be   & $0.30-0.52$ \cr
\hline
KamLAND    & 2001               & $^7$Be   & $0.30-0.52$ \cr
\hline
HELLAZ     & $>2002$            & $pp$     & $0.52-0.83$ \cr
\hline
HERON      & $>2000$            & $pp$     & $0.52-0.82$ \cr
\hline
\end{tabular}
\end{center} 
\vglue 0.5cm

The predictions for the rates in Tab.~1 reflect the relation between
the predominant neutrino fluxes that we presented in
eq.~(\ref{P_relation}), i.e. neutrinos with higher energies are in
general produced closer to the solar center and therefore more likely
to pass through a resonance and undergo flavor conversion.

As can be seen from Fig.~\ref{future} the suppression pattern of the
NSNI solution is clearly different from the one predicted by the small
angle MSW solution (c.f. Fig.~1 of Ref.~\cite{BKSSNO}). But there is
a striking similarity between the NSNI solution and the LMA solution,
including the preference for large $f_B$ (c.f. Fig.~7 of
Ref.~\cite{BKS}), the absence of a $^8$B spectral distortion and the
modest day night effect. Consequently, using solar neutrino data, it
will be difficult to distinguish the NSNI scenario from the LMA MSW
solution.  We note, however, that the KamLAND experiment will provide
an independent test of the oscillation parameters of the LMA MSW
solution by observing anti electron neutrinos from several nuclear
reactors around the Kamioka mine in Japan. Thus, if KamLAND would
indeed confirm the LMA MSW solution, then the NSNI solution discussed
in this paper will be irrelevant.

Since SNO~\cite{SNO} already started taking data and is expected to
have some results soon, let us consider some implication for this
experiment.  As we have pointed out one of the important features of
the NSNI conversion mechanism is the absence of any distortion in the
solar neutrino spectrum even though the averaged survival
probabilities of neutrinos from different nuclear reactions in the sun
are not equal.  Due to this feature, the following simple relation
between the SuperKamiokande solar neutrino event rate $R_{SK}$ and the
SNO CC event rate $R_{SNO}^{CC}$ (both normalized by the SSM
predictions) holds:
\beq \label{sksno}
R_{SK} = R_{SNO}^{CC} (1-r) + r f_B,  
\eeq
where $R_{SK}$ and $R_{SNO}^{CC}$ are defined exactly as in eqs.~(4)
and~(6) of Ref.~\cite{BKS99} and $r$ is given by
\beq 
r \equiv 
 \frac{\displaystyle \int dE_e R(E_e) \int dE_\nu 
 \phi^{^8\text{B}}(E_\nu)\sigma_{\nu_{\mu,\tau} e}(E_\nu,E_e) }
 {\displaystyle \int dE_e R(E_e) \int dE_\nu 
 \phi^{^8\text{B}}(E_\nu)\sigma_{\nu_ee}(E_\nu,E_e) }
\simeq \frac{1}{7} \,. 
\eeq
Here, $E_e$ and $E_\nu$ are the electron and neutrino energy,
respectively, $R(E_e)$ is the SuperKamiokande resolution and
efficiency function, $\phi^{^8\text{B}}$ is the $^8$B neutrino flux,
and $\sigma_{\nu_e e}$ and $\sigma_{\nu_{\mu,\tau} e}$ denote the
elastic scattering cross sections for $\nu_e \, e^- \to \nu_e \, e^-$
and $\nu_{\mu,\tau} \, e^- \to \nu_{\mu,\tau} \, e^-$, respectively.

We note that $R_{SK}$ and $R_{SNO}^{CC}$ are defined such that $R_{SK}
= R_{SNO}^{CC}$ in the absence of neutrino flavor transitions
including the case where $f_B \ne 1$.  (Strictly speaking, a slight
violation of the equality in eq.~(\ref{sksno}) could be induced by the
earth matter effect on these two experiments, since they are located
at somewhat different latitudes.)  Using the relation~(\ref{sksno}),
the true flux of the $^8$B neutrino flux $(\phi^{^8\text{B}})_{true} =
f_B (\phi^{^8\text{B}})_{SSM}$ could be precisely determined by
combining SuperKamiokande and SNO solar neutrino measurements, if the
solar neutrino problem is indeed due to NSNI.

Finally let us discuss shortly the possibility of testing the solution
studied in this paper by future long-baseline neutrino oscillation
experiments. Since only $\nu_e \to \nu_\tau$ transitions are viable,
an independent test would require a $\nu_\tau$ ($\bar\nu_\tau$)
appearance experiment using an intense beam of $\nu_e$
($\bar{\nu}_e$), which could be created at future neutrino factories
(see, e.g., Ref.~\cite{Geer}).

Assuming a constant density and using the approximation that $n_d
\simeq n_u \simeq 3n_e$ in the earth, the conversion probability for a
neutrino which travels a distance $L$ in the earth is given by:
\beq \label{oscprob}
P(\nu_e \to \nu_\tau;L) \simeq
 \frac{36\varepsilon^2}{36\varepsilon^2+(1-3\varepsilon')^2}
 \sin^2\left[ {1\over 2} 
 \sqrt{36\varepsilon^2+(1-3\varepsilon')^2} \sqrt{2} G_F n_e
 L \right] \,. 
\eeq
Numerically, the oscillation length in the earth matter can be
estimated to be
\beq \label{osclength}
L_{osc} \approx 8.1 \times 10^3 
\left[ \frac{2\ \mbox{mol/cc}}{n_e} \right]
\left[ \frac{1}{ \sqrt{36\varepsilon^2+(1-3\varepsilon')^2}} \right] \ 
 \mbox{km}.
\eeq
Using eqs.~(\ref{oscprob}) and~(\ref{osclength}) and the approximation
$n_e \sim 2$ mol/cc (which is valid close to the earth surface), we
find that, for the case of non-standard neutrino scattering off
$d$-quark, $P\sim$ few $\times 10^{-4}$ for K2K ($L = 250$~km) and
$P\sim$ few $\times 10^{-3}$ for MINOS ($L = 732$~km) for our best fit
parameters.  Similarly for $u$-quark, $P\sim$ few $ \times 10^{-5}$
for K2K and $P\sim$ few $\times 10^{-4}$ for MINOS for the best fit
parameters.  These estimates imply that it would be hard but not
impossible, at least for the case of scattering off $d$-quarks, to
obtain some signal of $\nu_e \to \nu_\tau$ conversions due to NSNI
interactions by using an intense $\nu_e$ beam which can be created by
a muon storage ring~\cite{Geer}.

\section{Conclusions}

According to our $\chi^2$ analysis \NSNI\ (NSNI) can provide a good
fit to the solar neutrino data provided that there are rather large
non-universal FDNI (of order $0.5 \, G_F$) and small FCNI (of order
$10^{-2}-10^{-3} \, G_F$). The fit to the observed total rate,
day-night asymmetry, seasonal variation and spectrum distortion of the
recoil electron spectrum is comparable in quality to the one for
standard neutrino oscillations.

From the model-independent analysis we learn that NSNI induced by the
exchange of heavy bosons cannot provide large enough $\nu_e \to
\nu_\mu$ transitions, while $\nu_e - \nu_\tau$ FCNI in principle could
be sufficiently strong. However, the current bounds will be improved
by the up-coming $B$-factories, providing an independent test of the
NSNI solution.  The required large non-universal FDNI (for $\nu_e$
transitions into both $\nu_\mu$ and $\nu_\tau$) can be ruled out by
the upper bounds on lepton universality, unless they are induced by an
intermediate doublet of $SU(2)_L$ (a scalar or a vector boson) or by a
neutral vector singlet. For $\nu_e \to \nu_\mu$ there exists a bound
due to the limit on compositeness in this case, but for $\nu_e \to
\nu_\tau$ there is no significant constraint at present.

Generically only very few models can fulfill the requirements needed
for the solution discussed in this paper: massless neutrinos, small
FCNI and relatively large non-universal FDNI. As for the vector bosons
the most attractive scenario is to evoke an additional
$U(1)_{B-3L_\tau}$ gauge symmetry (where $B$ is the baryon number and
$L_\tau$ denotes the tau lepton number), which would introduce an
additional vector singlet that only couples to the third generation
leptons and quarks~\cite{MR}. Among the attractive theories beyond the
standard model where neutrinos are naturally massless as a result of a
protecting symmetry, are supersymmetric $SU(5)$ models~\cite{FCSU5}
that conserve $B-L$, and theories with an extended gauge structure
such as $SU(3)_C \otimes SU(3)_L \otimes U(1)_N$ models~\cite{331},
where a chiral symmetry prevents the neutrino from getting a mass.
These particular models, however, do not contribute significantly to
the specific interactions we are interested in this paper.  $SU(5)$
models have negligible NSNI since they are mediated by vector bosons
which have masses at the GUT scale.  $SU(3)_C \otimes SU(3)_L \otimes
U(1)_N$ models can provide large $\epsilon_e$ and $\epsilon'_e$, but
these models do not induce NSNI with quarks. From eq.~(\ref{res}) it
follows that no resonant conversion can occur in this case.

Therefore we conclude that the best candidate for the scenario we
studied are supersymmetric models with broken $R$-parity, where the
relevant NSNI are mediated by a scalar doublet, namely the
``left-handed'' bottom squark.  Although in this model neutrino masses
are not naturally protected from acquiring a mass, one may either
evoke an additional symmetry or assume that non-zero neutrino masses
are not in a range that would spoil the solution in terms of the
non-standard neutrino oscillations we have studied in this paper.

Even though we consider the conventional oscillation mechanisms as the
most plausible solutions to the solar neutrino problem, it is
important to realize that in general New Physics in the neutrino
sector include neutrino masses and mixing, as well as new neutrino
interactions. While it is difficult to explain the atmospheric
neutrino problem~\cite{AN} and the LSND anomalies~\cite{LSND} by
NSNI~\cite{BG,BGP}, we have shown in this paper that a solution of the
solar neutrino problem in terms of NSNI is still viable. The ultimate
goal is of course a direct experimental test of this solution. The
upcoming solar neutrino experiments will provide a lot of new
information which hopefully will reveal the true nature of the solar
neutrino problem.

\acknowledgments 

This work was partially supported by Funda\c{c}\~ao de Amparo \`a
Pesquisa do Estado de S\~ao Paulo (FAPESP), Conselho Nacional de
Desenvolvimento Cient\'\i fico e Tecnolog\'ogico (CNPq) and NSF grant
PHY-9605140.  We thank Y.  Grossman, Y. Nir and C. Pe\~na-Garay for
helpful discussions.  One of us (MMG) would like to thank the Physics
Department, University of Wisconsin at Madison, where part of this
work was realized, for the hospitality.




\newpage


\newpage


\vglue 0.5cm


\newpage


\newpage 
\vglue 0.5cm 



\begin{thebibliography}{99}

\bibitem{HS} 
T.B. Cleveland {\it et al.} (Homestake Collaboration),
Astrophys. J. {\bf 496}, 505 (1998).

\bibitem{GALLEX}  
W. Hampel {\it et al.} (GALLEX Collaboration),
\plb{447}, 127 (1999).

\bibitem{SAGE} 
J.N. Abdurashitov {\it et al.} (SAGE Collaboration), 
\prc{60}, 055801 (1999).

\bibitem{KK} 
Y. Fukuda  {\it et al.} (Kamiokande Collaboration), 
\prl{77}, 1683 (1996).

\bibitem{SK} 
Y. Fukuda et al. (SuperKamiokande Collaboration), 
\prl{82}, 1810 (1999).

\bibitem{JNB} 
J.N. Bahcall, 
Neutrino Astrophysics, Cambridge University Press, 1989. 

\bibitem{BP95}  
J.N. Bahcall and M.H. Pinsonneault, 
\rmp{67}, 781 (1995).

\bibitem{BPBC}
J.N. Bahcall, M.H. Pinsonneault, S. Basu and J. Christensen-Dalsgaard, \\
\prl{78}, 171 (1997).

\bibitem{BBP} 
J.N. Bahcall, S. Basu and M.H. Pinsonneault, 
\plb{433}, 1 (1998); see also 
J.N. Bahcall's home page, {\tt http://www.sns.ias.edu/$^\sim\!$jnb}.

\bibitem{olderSSM} 
See also, 
J.N. Bahcall and R.K. Ulrich, \rmp{60}, 297 (1988);\\
J.N. Bahcall and M.H. Pinsonneault, \rmp{64}, 885 (1992);\\
S. Turck-Chieze {\it et al.}, Astrophys. J. {\bf 335}, 415 (1988);\\
S. Turck-Chieze and I. Lopes, Astrophys. J. {\bf 408}, 347 (1993); \\
V. Castellani, S. Degl'Innocenti and G. Fiorentini, 
Astron. Astrophys. {\bf 271}, 601 (1993); \\
V. Castellani {\it et al.}, \plb{324}, 425 (1994). 

\bibitem{Adel}  
E.G. Adelberger et al., 
\rmp{70}, 1265 (1998).

\bibitem{30years} 
For a detailed list of the references on these solutions see: \\  
``Solar Neutrinos: The First Thirty Years'', 
ed. by R. Davis Jr. {\sl et al.}, \\
Frontiers in Physics, Vol. 92, 
Addison-Wesley, 1994. 

\bibitem{sterile}   
See for example: \\
D.O. Caldwell and R.N. Mohapatra, \prd{48}, 3259 (1993); \\
J.\ T. Peltoniemi, D.\ Tommasini, and J.W.F. Valle, 
Phys. Lett. B {\bf298}, 383 (1993); \\  
J.T. Peltoniemi and J.W.F. Valle, \npb{406}, 409 (1993); \\  
G. Dvali and Y. Nir, \JHEP{9810}, 014 (1998) [{\tt hep-ph/9810257}]; \\  
G. M. Fuller, J. R. Primack and Y.-Z. Qian, 
 Phys. Rev. D{\bf 52}, 1288 (1995); \\  
J. J. Gomez-Cadenas and M. C. Gonzalez-Garcia, 
 Zeit. fur Physik {\bf C71}, 443 (1996); \\  
N. Okada and O. Yasuda, Int. J. Mod. Phys. A {\bf 12}, 3669 (1997), 
 and references therein. 

\bibitem{vacuum} 
V. N. Gribov and B. M. Pontecorvo, 
\plb{28}, 493 (1969).

\bibitem{wolf} 
L. Wolfenstein, 
Phys. Rev. {\bf D17}, 2369 (1978).

\bibitem{MS} 
S.P. Mikheyev and A. Yu. Smirnov,  
\sjnp{42}, 913 (1985); \\
Nuovo Cimento {\bf C9}, 17 (1986).

\bibitem{BKS} 
J.N. Bahcall, P.I. Krastev and A.Yu. Smirnov, \\
\prd{58}, 096016 (1998) {[{\tt hep-ph/9807216}]};\\ 
\prd{60}, 093001 (1999) {[{\tt hep-ph/9905220}]}.

\bibitem{GHPV} 
M.C. Gonzalez-Garcia, P.C. de Holanda, C. Pe\~na-Garay and
J.W.F. Valle, \\
{\tt hep-ph/9906469}, \npb, in press. 

\bibitem{FLMP} 
G.L. Fogli, E. Lisi, D. Montanino and A. Palazzo, 
{\tt hep-ph/9912231}. 


\bibitem{GMP} 
M.M. Guzzo, A. Masiero and S.T. Petcov, 
\plb{260}, 154 (1991).

\bibitem{BPW} 
V. Barger, R.J.N. Phillips and K. Whisnant, 
\prd{44}, 1629 (1991).

\bibitem{ER} 
E. Roulet, 
\prd{44}, 935 (1991). 

\bibitem{DR}
S. Degl'Innocenti and B. Ricci, \mpla{8}, 471 (1993).

\bibitem{FL94}
G.L. Fogli and E. Lisi, 
Astroparticle Phys. {\bf 2}, 91 (1994).

\bibitem{KB} 
P.I. Krastev and J.N. Bahcall, 
``FCNC solutions to the solar neutrino problem'', \\ 
{\tt hep-ph/9703267}.

\bibitem{Bergmann}
S. Bergmann, 
\npb{515}, 363 (1998) [{\tt hep-ph/9707398}].

\bibitem{Valle87} 
Resonant neutrino conversion induced by non-orthogonal massless
neutrinos was first discussed by J.W.F. Valle in \plb{199}, 432
(1987). \\
%
However this mechanism can not induce a large effect on the solar
neutrinos due to the stringent constraints on the model parameters.
See also Refs.~\cite{Z-FCNC,BergmannKagan}.

\bibitem{SUSYwoR}
C.S. Aulakh and N.R. Mohapatra, \plb{119}, 136 (1983); \\
F. Zwirner, \plb{132}, 103 (1983); \\
L.J. Hall and M. Suzuki, \npb{231}, 419 (1984); \\
J. Ellis {\it et al.}, \plb{150}, 14 (1985); \\
G.G. Ross and J.W.F. Valle,  \plb{151}, 375 (1985); \\
R. Barbieri and A. Masiero, \plb{267}, 679 (1986).

\bibitem{BGN}       
S. Bergmann, Y. Grossman and E. Nardi, 
\prd{60}, 093008 (1999) \\
{[{\tt hep-ph/9903517}]}.
 
\bibitem{Z-FCNC}    
P. Langacker and D. London, \prd{38}, 886 (1988); 
{\it ibid} {\bf 38}, 907 (1988); \\
H. Nunokawa, Y.-Z. Qian, A. Rossi and J.W.F. Valle, \prd{54}, 4356 (1996) \\ 
{[{\tt hep-ph/9605301}]}, and references therein.

\bibitem{BergmannKagan} 
S. Bergmann and A. Kagan, 
\npb{538}, 368 (1999) [{\tt hep-ph/9803305}].

\bibitem{eartheffect}
S.P. Mikheyev and A.Yu. Smirnov, 
Sov. Phys. Usp. {\bf 30}, 759 (1987). 

\bibitem{Freund}
M. Freund and T. Ohlsson, {\tt hep-ph/9909501}.

\bibitem{Akhmedov}
E.Kh. Akhmedov, 
\npb{538}, 25 (1999).

\bibitem{FLM}
G.L. Fogli, E. Lisi and D. Montanino, 
\prd{49}, 3226 (1994).

\bibitem{FL} 
G.L. Fogli and E. Lisi,
Astropart. Phys. {\bf 3}, 185 (1995).

\bibitem{Suzuki} 
Y. Suzuki,
``Solar Neutrinos'', talk given at the Lepton Photon Conference, 1999.

\bibitem{JNB2} 
See Fig. 8.2 of Ref.~\cite{JNB} for the recoil electron spectra shape 
from $^8$B neutrino due to $\nu_e e^- \to \nu_e e^-$ 
and $\nu_{\mu,\tau} e^- \to \nu_{\mu,\tau} e^-$ 
scattering. 

\bibitem{seasonal}
P.C.de Holanda, C.Pe\~na-Garay, M.C.Gonzalez-Garcia and J.W.F.Valle, 
\prd{60} (1999) 093010.

\bibitem{BG} 
S. Bergmann and Y. Grossman, 
\prd{59}, 093005 (1999) 
[{\tt hep-ph/9809524}].

\bibitem{BGP} 
S. Bergmann, Y. Grossman and D.M. Pierce, 
\prd{61}, 53005 (2000) \\
{[{\tt hep-ph/9909390}]}.

\bibitem{LSND} 
C. Athanassopoulos {\it et al.} (LSND Collaboration), 
\prl{77}, 3082 (1996); \\
\prl{81}, 1774 (1998).

\bibitem{AN}        
For recent analysis, see {\it e.g.}: \\
M.C. Gonzalez-Garcia, {\it  et al.}, \prd{58}, 033004 (1998); \\
M.C. Gonzalez-Garcia, H. Nunokawa, O.L.G. Peres and J.W.F. Valle, \\
\npb{543}, 3 (1999) [{\tt hep-ph/9807305}];\\
N. Fornengo, M.C. Gonzalez-Garcia and J.W.F. Valle, {\tt hep-ph/0002147}. 

\bibitem{LNV}
S. Bergmann, H.V. Klapdor--Kleingrothaus and H. P\"as, {\tt hep-ph/0004048}.

\bibitem{Bliss}
W. Bliss {\it et al.} (CLEO Collaboration), 
\prd{57}, 5903 (1998) [{\tt hep-ex/9712010}]. 

\bibitem{PDG}       
Particle Data Group: C. Caso {\it et al.}, \epjc{3}, 1 (1998); see also: \\
{\tt http://pdg.lbl.gov}\,.

\bibitem{Don}
J.F. Donoghue, E. Golowich and B.R. Holstein, 
{\it Dynamics of the Standard Model}, \\
Cambridge University Press, 1992.

\bibitem{Pich} 
A. Pich, Review talk at Lepton Photon 99, Stanford University, 
August, 1999. \\
See also: {\tt hep-ph/9711279, hep-ph/9802257}. 

\bibitem{Abe} 
F. Abe {\it et al.} (CDF Collaboration), \prl{79}, 2198 (1997).

\bibitem{GGN} 
M.C. Gonzalez-Garcia, A. Gusso and S.F. Novaes, 
J. Phys. {\bf G}, 2213 (1998) \\ {[{\tt hep-ph/9802254}]}.

\bibitem{obl}
M.E. Peskin and T. Takeuchi, 
\prl{65}, {964} (1990); \\
\prd{46}, 381 (1992).

\bibitem{BaBar}
{\it The BaBar physics book} (SLAC--R--504), Chapter 12.2.6, 
available at: \\
{\tt http://www.slac.stanford.edu/pubs/slacreports/slac-r-504.html}\,.

\bibitem{SNO}
A. B. McDonald, for the SNO collaboration, 
\npb{77} (Proc. Suppl.), 43 (1999). 

\bibitem{BOREXINO}
L. Oberauer, 
\npbps{77}, 48 (1999). 

\bibitem{KamLAND}
A. Suzuki, 
\npbps{77}, 171 (1999). 

\bibitem{HELLAZ}
A. De Bellefon for HELLAZ collaboration, 
\npbps{70}, 386 (1999). 

\bibitem{HERON}
R. E. Lanou,  
in Proceedings of the 8th International Workshop on Neutrino Telescopes 
(Venice, Italy, 1999), ed. by M. Baldo Ceolin, Vol. I, page 139. 

\bibitem{BKSSNO} 
J.N. Bahcall, P.I. Krastev and A.Yu. Smirnov, 
{{\tt hep-ph/0002293}}.

\bibitem{BKS99} 
J.N. Bahcall, P.I. Krastev and A.Yu. Smirnov, 
\plb{477}, 401 (2000) [{\tt hep-ph/9911248}].

\bibitem{Geer} 
S. Geer, \prd{57}, 6989 (1998).

\bibitem{MR}
E. Ma and D. P. Roy, \prd{58}, 095005 (1998) [{\tt hep-ph/9806210}], \\
E. Ma, \plb{433}, 74 (1998) [{\tt hep-ph/9709474}].

\bibitem{FCSU5} 
L.J. Hall, V.A. Kostelecky, S. Raby, \npb{B267} 415 (1986); \\ 
Y. Okada, {\tt hep-ph/9809297}.

\bibitem{331} 
F. Pisano and V. Pleitez, \prd{46}, 410 (1992); \\
P. Frampton, \prl{69}, 2889 (1992).

\end{thebibliography}
\end{document}